\documentclass[12pt]{iopart}
\expandafter\let\csname equation*\endcsname\relax
\expandafter\let\csname endequation*\endcsname\relax
\usepackage{iopams}  
\usepackage{amsmath,amssymb}
\usepackage{graphics}
\usepackage{graphicx}
\usepackage{epsfig}
\usepackage{float}
\usepackage[usenames]{color}
\usepackage{subfigure}

\newcommand{\boldnabla}{\mbox{\boldmath$\nabla$}}

\def\mc{\mathcal}

\newcommand {\apgt} {\ {\raise-.5ex\hbox{$\buildrel>\over\sim$}}\ }
\newcommand {\aplt} {\ {\raise-.5ex\hbox{$\buildrel<\over\sim$}}\ }

\newcommand{\longtitle}[1]{%
  \ifodd\value{page}%
    \protect\parbox{0.97\linewidth}{#1}\hfill%
  \else%
    \hfill\protect\parbox{0.97\linewidth}{#1}%
  \fi%
}

\begin{document}

\title{The correlation between the Nernst effect and fluctuation diamagnetism in strongly
 fluctuating superconductors}

\author{Kingshuk Sarkar$^1$ , Sumilan Banerjee$^{1}$, Subroto Mukerjee$^{1,2}$, T. V. Ramakrishnan$^{1,3}$ }
\address{$^1$ Department of Physics, Indian Institute of Science, Bangalore 560 012, India\\
$^2$ Centre for Quantum Information and Quantum Computing, Indian Institute of Science,
Bangalore 560 012, India\\
$^3$ Department of Physics, Banaras Hindu University, Varanasi 221005, India}

\hspace{1cm}E-mail: kingshuk@physics.iisc.ernet.in, sumilan@physics.iisc.ernet.in,\\
smukerjee@physics.iisc.ernet.in, tvrama2002@yahoo.co.in

\hspace{1cm}Keywords:  Nernst effect, Transverse thermoelectric transport coefficient, Diamagnetism,
Correlation, Superconducting fluctuations, Cuprates

\begin{abstract}

 We study the Nernst effect in fluctuating superconductors by calculating the transport coefficient $\alpha_{xy}$ in a phenomenological model where relative importance of phase and amplitude fluctuations of the order parameter is tuned continuously to smoothly evolve from an effective XY model to more conventional Ginzburg-Landau description. To connect with a concrete experimental realization we choose the model parameters appropriate for cuprate superconductors and calculate $\alpha_{xy}$ and the magnetization ${\bf M}$ over the entire range of experimentally accessible values of field, temperature and doping. We argue that $\alpha_{xy}$ and ${\bf M}$ are both determined by the equilibrium properties of the superconducting fluctuations (and not their dynamics) despite the former being a transport quantity. Thus, the experimentally observed correlation between the Nernst signal and the magnetization arises primarily from the correlation between $\alpha_{xy}$ and ${\bf M}$.  Further, there exists a dimensionless ratio ${\bf M}/(T \alpha_{xy})$ that quantifies this correlation. We calculate, for the first time, this ratio over the entire phase diagram of the cuprates and find it agrees with previous results obtained in specific parts of the phase diagram. We conclude that that there appears to be no sharp distinction between the regimes dominated by phase fluctuations and Gaussian fluctuations for this ratio in contrast to $\alpha_{xy}$ and ${\bf M}$ individually. The utility of this ratio is that it can be used to determine the extent to which superconducting fluctuations contribute to the Nernst effect in different parts of the phase diagram given the measured values of magnetization. 
\end{abstract}

\title[\longtitle{The correlation between the Nernst effect and fluctuation diamagnetism in strongly
 fluctuating superconductors}]
\maketitle

\section{Introduction}

The Nernst effect is the phenomenon of the production of an electric field ${\bf E}$ in a direction perpendicular to an applied temperature gradient ${\bf \nabla} T$ under conditions of zero electrical current flow. This is possible only when time reversal symmetry is broken and thus in the most common setting the sample is placed in an external magnetic field ${\bf B}$. The Nernst effect is particularly pronounced in type II superconducting systems~\cite{Palstra_1990, Ong_2000, Ong_2006, Pourret_2006}. Such systems possess mobile vortices for certain ranges of values of applied magnetic field and temperature. These vortices can move under the influence of a temperature gradient inducing a transverse electric field through phase slips. The vortices possess entropy which causes them to move opposite to the direction of an applied temperature gradient. However, since they carry no charge they do not produce an electric current giving rise to the Nernst effect. The Nernst signal is proportional to the vortex entropy. In contrast, for systems in which the elementary mobile degrees of freedom are charged quasiparticles, the condition of zero electrical current implies an equal and opposite flux of particles along and against the temperature gradient. The particles moving in the two opposite directions carry different amounts of entropy giving rise to a heat current. However, if they are scattered in the same way, the transverse electric fields induced by them cancel in the presence of a magnetic field giving rise to a zero Nernst signal. This is known as the Sondheimer cancellation~\cite{Sondheimer_1948}. The Nernst effect in quasiparticle systems is thus typically produced by energy dependent scattering or amibipolarity of the carriers and is generally not as strong as in superconductors. The Nernst effect has also been observed in heavy fermion systems~\cite{Bel_2004,Luo_2016}.

The above discussion would suggest that a pronounced Nernst signal in a superconductor is an indicator of mobile vortices. However, the Nernst effect has been observed in the cuprates at temperatures well above the transition temperature $T_c$~\cite{Ong_2000,Ong_2006}. A description of the system in terms of distinct non-overlapping vortices is not always possible at such high temperatures. In overdoped cuprates, it has been argued that the Nernst effect is most effectively described in terms of Gaussian fluctuations of the superconducting order parameter rather than distinct mobile vortices~\cite{Ussishkin}. Calculations of the Nernst coefficient in this regime at small magnetic fields produce a good match to experimental data at low fields. At high fields and low temperatures, the Gaussian theory is not applicable. Nevertheless, a description of the system in terms of a Ginzburg-Landau theory of superconducting fluctuations with appropriate dynamics produces a good match to experimental data~\cite{Mukerjee}. Other works along similar lines include a calculation based on self-consistent Gaussian approximation using Landau level basis at low temperature and finite fields~\cite{Rosenstein_2009,Tinh_2014} and a Coulomb gas model of vortices with the core energy related to the Nernst effect and diamagnetism~\cite{Orgad1_2014,Orgad2_2014,Orgad_2015}.

In the underdoped region, fluctuations are expected to be much stronger yielding a large region of temperature with dominant fluctuations in the phase of the order parameter with a largely uniform amplitude. A description of the system in terms of mobile vortices is a good one in this regime and a calculation of the Nernst effect based on a classical $XY$ model has been performed yielding a good match to experimental data~\cite{Podolsky}.  A systematic interpolation between these two regimes as a function of doping, temperature and magnetic field for the Nernst has been lacking, primarily due to the absence of a common theory of superconducting fluctuations across the entire superconducting phase diagram. In this paper, we address this lacuna in the literature by employing a phenomenological Ginzburg-Landau-type functional developed by two of us~\cite{Banerjee_1, Banerjee_2}. Calculations based on this functional have provided good agreement with experimental measurements of different quantities such as the specific heat, superfluid density, photoemission and the superconducting dome across the entire range of doping and temperature of the cuprate phase diagram. This functional has also recently been employed by us to obtain a fairly good agreement with measurements of fluctuation diamagnetism in the cuprates~\cite{Sarkar_2016}. 

The measured Nernst effect in different parts of the cuprate phase diagram has been variously attributed to Gaussian fluctuations~\cite{Ussishkin}, phase fluctuations~\cite{Podolsky} and quasiparticles~\cite{Sachdev_2010}. In several instances there is no consensus on exactly which mechanism is responsible for the observed signal in the same part of the phase diagram~\cite{Ong_2001,Ong_2006,Taillefer_2010,Varlamov_2011} also complicated by the observation of competing orders. In this work we calculate the coefficient $\alpha_{xy}$, called the off-diagonal Peltier coefficient and sometimes the Ettingshausen coefficient, from a model of superconducting fluctuations. In the limit of strong particle-hole symmetry, as seen for many superconductors, the Nernst coefficient $\nu = \frac{1}{H} \frac{\alpha_{xy}}{\sigma_{xx}},$ where $H$ is the magnetic field and $\sigma_{xx}$, the magnetoconductivity. We show that in a model of superconducting fluctuations, $\alpha_{xy}$, despite being a transport quantity,  is expected to be naturally related to equilibrium quantities. This is due to the fact that $\alpha_{xy}$ is determined by the strength of the superconducting fluctuations as opposed to their dynamics (as we explain later), which is also responsible for equilibrium phenomena. On the other hand, $\nu$ and $\sigma_{xx}$ are given by the dynamics of the fluctuations. In particular, we argue that $\alpha_{xy}$ is naturally related to the magnetization ${\bf M}$ through a dimensionless ratio ${\bf M}/(T \alpha_{xy})$, which is a function of doping, temperature and magnetic field. Experimentally, in hole-doped cuprate superconductors above the superconducting transition temperature $T_c$ in the pseudogap regime a large diamagnetic response has  been observed concurrently with a large Nernst signal over a wide range of temperatures~\cite{Ong_2005,Li_2007,Li_2010,Xiao}. A connection between $\alpha_{xy}$ and ${\bf M}$ via the ratio ${\bf M}/(T \alpha_{xy})$ has also been proposed theoretically in the $XY$ and Gaussian fluctuation dominated regime of the cuprate phase diagram~\cite{Ussishkin,Podolsky,Raghu,Tinh_2014} and found to be consistent with experimental observations. In most superconductors, including the cuprates, superconducting fluctuations are the main source of any large observed diamagnetic signal. Thus, a concurrent measurement of $\alpha_{xy}$ along with a comparison to our calculated ratio of ${\bf M}/(T \alpha_{xy})$ can provide an indication of whether the observed Nernst signal is also due to superconducting fluctuations. We illustrate this by performing our calculations on our phenomenological model of superconducting fluctuations for the cuprates, mentioned in the previous paragraph.

The paper is organized as follows: In section 2, we discuss the model we study and various details concerning the form of the currents and transport coefficients obtained from it. Section 3 contains a discussion of the methodology and a description of the details of our numerical simulations. We present the results of our simulations in section 4 and comment on the important features seen in the data. Finally in section 5, we discuss the novel findings of our calculations and also their relation to previous theoretical and experimental work. Additionally, there are three appendices which discuss technical details pertinent to the calculations and results discussed in the main text.

 \section{Model}\label{SC.Model}

To study transport properties due to superconducting fluctuations we implement ``model A'' dynamics for a complex superconducting order parameter $\Psi(r,t)$ given by the stochastic equation
\begin{equation}\label{Eq.TDGL}
\tau D_t\Psi(r,t)=-\frac{\delta F\lbrace \Psi, \Psi^{*}\rbrace}{\delta \Psi^{*}(r,t)}+\eta.
\end{equation}
$F\lbrace \Psi, \Psi^{*}\rbrace$ is a free energy functional.
In order to be able to introduce electromagnetic fields, we define a covariant time derivative $D_t=(\frac{\partial}{\partial t}+i\frac{2\pi}{\Phi_0}\Phi)$ and a covariant spatial derivative ${\bf D}=\nabla-i\frac{2\pi}{\Phi_0}{\bf A}$. ${\bf A}(r,t)$ and ${\Phi}(r,t)$ are the magnetic vector and scalar potential respectively while $\Phi_0=\frac{h}{e^{*}}$ is the flux quantum. The free energy functional is assumed to contain an energy cost for spatial inhomogeneities of the order parameter through the appearance of terms involving the covariant spatial derivative. The specific model we study is defined on a lattice, where the spatial derivative has to be appropriately discretized as we discuss later. The time scale $\tau$, which provides the characteristic temporal response scale of the order parameter dynamics, can in general be complex. However it is required to be real under the requirement that the equation of motion for $\Psi^{*}$ be the same as for $\Psi$ under the simultaneous transformation of complex conjugation ($\Psi\rightarrow \Psi^{*}$) and magnetic field inversion ($H\rightarrow -H$) (particle-hole symmetry). Evidence of particle-hole symmetry in the form of no appreciable Hall or Seebeck effect is seen in the experimentally accessible regime of the superconductors we study here and thus we take $\tau$ to be real in our calculations. 
The thermal fluctuations are introduced through  $\eta({\bf r},t)$  with the Gaussian white noise correlator
\begin{equation}\label{Eq.noise}
\langle \eta^{*}({\bf{r}},t)\eta({\bf{r}}^{\prime},t^{\prime}) \rangle =2k_BT\tau\delta({\bf{r}}-{\bf{r^{\prime}}})\delta(t-t^{\prime})
\end{equation}
Further, the magnetic field (${\bf H}=\nabla\times{\bf A}$) is assumed to be uniform and not fluctuating due to a large ratio ($\kappa$) between the London penetration depth ($\lambda$) and the coherence length ($\xi$) for the strong type-II superconductors we study. Cuprate and iron-based superconductors are examples of these. 

The dynamical model Eq.~\ref{Eq.TDGL} is the simplest one which yields an equilibrium state in the absence of driving potentials. It can be derived microscopically within BCS theory above and close to the transition temperature $T_c$. However, it has been used phenomenologically to study transport previously in situations, where the microscopic theory is not known, such as for the cuprates~\cite{Ussishkin,Mukerjee,Podolsky}. We employ the model in a similar spirit here.

\subsection{Heat and electrical transport coefficients }\label{sec.mag} 
The model described by Eq.~\ref{Eq.TDGL} has no conservation laws and thus currents cannot be defined in terms of continuity equations. Nevertheless, they can be defined by appealing to the microscopics of the full system and then identifying the degrees of freedom that contribute to the superconductivity. The expression for the charge current density obtained this way is~\cite{Ussishkin, Caroli,UD_1991,A_Schmid}
 \begin{equation}\label{Eq.Charge_Current}
 {\bf J}^{ \rm e}_{\rm tot}=-\frac{\delta F}{\delta A}
 \end{equation}
 An expression can also be obtained for the heat current density ${\bf J}^{\rm Q}$ along similar lines but it cannot be written as compactly as the one for the charge current density~\cite{Caroli, A_Schmid}. We provide the exact expression for the heat current for the model we study in the next subsection. For the present discussion, we only require that ${\bf J}^{\rm Q}$ exists. 
In the presence of a magnetic field, these current densities are sums of transport and magnetization current densities ~\cite{Cooper_1997}.
\begin{eqnarray}
{\bf J}^{\rm {e}}_{\rm tot}({\bf r}) & = & {\bf J}^{\rm{e}}_{\rm tr}({\bf r}) + {\bf J}^{\rm {e}}_{\rm mag}({\bf r}) \\
{\bf J}^{\rm{Q}}_{\rm tot}({\bf r}) & = & {\bf J}^{\rm{Q}}_{\rm tr}({\bf r}) + {\bf J}^{\rm{Q}}_{\rm mag}({\bf r}) ~,\nonumber
\end{eqnarray}
where tr and mag stand for transport and magnetization respectively.

The transport coefficients we calculate are described only by the transport parts of the current densities, to obtain which the magnetization parts need to be subtracted from the total current densities. We detail the steps to do this in~\ref{app.Current_densities} which follows the discussion of ref.~\cite{Cooper_1997}.

The transport current densities can be related to an applied temperature gradient $\nabla T$ and electric field $\bf{E}$ in linear response as
\begin{center}
$\begin{pmatrix} {\bf J}_{\mathrm{tr}}^e \\ {\bf J}_{\mathrm{tr}}^Q \end{pmatrix} = \begin{pmatrix} \hat{\sigma} & \hat{\alpha} \\ \ \hat{\tilde{\alpha}} & \hat{\kappa} \end{pmatrix}   \begin{pmatrix} \bf{E} \\ -\nabla{T}, \end{pmatrix}$
\end{center}
where $\hat{\sigma}$, $\hat{\alpha}$, $\hat{\tilde{\alpha}}$, $\hat{\kappa}$ are the electrical, thermoelectric, electro-thermal and thermal conductivity tensors respectively and are independent of the gradients in linear response. On general grounds it can be shown that $\sigma_{xy}(H)=-\sigma_{yx}(H)$ and $\alpha_{xy}(H)=-\alpha_{yx}(H)$. The Nernst co-efficient ($\mathrm{\nu}$) under the condition $\rm{J^e_{tr}=0}$ is given by~\cite{Ussishkin,UD_1991}
\begin{equation}
\nu=\dfrac{E_y}{H\nabla_x T}=\frac{1}{H}\frac{\alpha_{xy}\sigma_{xx}-\sigma_{xy}\alpha_{xx}}{\sigma_{xx}^2+\sigma_{xy}^2}
\end{equation}
For systems with particle-hole symmetry $\alpha_{xx}$ and $\sigma_{xy}$ are zero and thus
\begin{equation}
\nu=\frac{\alpha_{xy}}{H\sigma_{xx}}
\end{equation}
Further, the Onsager relation gives $\hat{\tilde{\alpha}}=T\hat{\alpha}$~\cite{Cooper_1997}.

\subsection{Dimensional analysis of the transport coefficients}

Eq.~\ref{Eq.TDGL} can be written in terms of dimensionless parameters as follows. We assume that there are basic scales, $x_0$, $T_0$ and $\Psi_0$ for the spatial coordinate, temperature and the order parameter arising in the equilibrium state of the system. We can then define ${\bf r'}$, $T'$ and $\Psi'$, which are the dimensionless spatial coordinate, temperature and order parameter respectively by scaling by the quantities $x_0$, $T_0$ and $\Psi_0$. Eqs.~\ref{Eq.TDGL} and~\ref{Eq.noise} can now be cast in dimensionless 
form in terms of these quantities as 
\begin{equation}\label{Eq.TDGLnodim}
D_{t'}\Psi'=-\frac{\delta F'}{\delta \Psi'^{*}}+\eta'
\end{equation}
and
\begin{equation}\label{Eq.noisenodim}
\langle \left( \eta'({\bf r'}_1,t'_1) \right)^*\eta'({\bf r'}_2,t'_2) \rangle =2T'\delta({\bf r'}_1-{\bf r'}_2)\delta(t'_1-t'_2),
\end{equation}
where $t'$, $F'$ and $\eta'$ are the dimensionless values of the time, free energy density and noise. This is possible only if their basic scales are $t_0=\frac{\tau (\Psi_0)^2(x_0)^d}{k_BT_0}$, $F_0=\frac{k_BT_0}{(x_0)^d}$ and $\eta_0=\frac{\Psi_0 (x_0)^d}{k_BT_0}$ respectively, where $d$ is the number of spatial dimensions. Additionally, the basic scale of the magnetic flux is $\Phi_0$, which from gauge invariance implies that the basic scales of the electric potential $V$ and electrical current density ${\bf J}^e$ are $V_0=\frac{\Phi_0}{t_0}$ and $J^e_0=\frac{k_BT}{(x_0)^{d-1}\Phi_0}$. Thus, the basic scales of the coefficients $\hat{\sigma}$ and $\hat{\alpha}$ are $\frac{J^e_0 x_0}{V_0}$ and $\frac{J^e_0 x_0}{T_0}$. The dimensionless quantities $\hat{\sigma}$ and $\hat{\alpha}$ can be calculated from Eqns.~\ref{Eq.TDGLnodim} and~\ref{Eq.noisenodim} using the dimensionless form of ${\bf J}^e$. These can then be multiplied by appropriate basic scales to get their correct dimensional values. 

From the above discussion, it can be seen that while $\hat{\sigma}$ is proportional to the relaxation time $\tau$, $\hat{\alpha}$ is independent of it. Thus, the Nernst signal is inversely proportional to $\tau$ in our model. $\alpha$ depends only on the parameters of $F$ which also determine thermal equilibrium properties of the system. In particular, the ratio $\frac{|{\bf M}|}{T \alpha}$ is dimensionless, where ${\bf M}$ is the magnetization, suggesting a possible relationship between ${\bf M}$ and $\alpha$. In this work, we thus assert that the most meaningful comparison of fluctuation diamagnetism with the Nernst effect is a comparison of $\alpha_{xy}$ and ${\bf M}$. 

It has been shown that for a fluctuating 2D superconductor in the limit of Gaussian superconducting fluctuations and low magnetic fields $\frac{|{\bf M}|}{T \alpha_{xy}}=2$~\cite{Ussishkin}. Interestingly, in the complementary limit of very strong fluctuations with temperature much higher than $T_c$ and weak fields, the same ratio is obtained~\cite{Podolsky}. In this work, we calculate this ratio without restricting ourselves to the above limits and show that it in general deviates from the value of 2.

 \subsection{The free energy functional}
The free energy functional we use describes superconductivity on a two dimensional lattice~\cite{Banerjee_1}. It has a Ginzburg-Landau form with parameters chosen to reproduce experimental observations for the cuprates. In particular, it has been employed to successfully reproduce experimental measurements of the specific heat, superfluid density, superconducting dome and fluctuation diamagnetism~\cite{Banerjee_1,Sarkar_2016} . Coupling nodal quasiparticles to the fluctuations produces Fermi arcs~\cite{Banerjee_2}. The functional essentially describes the cuprates as highly anisotropic layered materials with weakly coupled stacks of $\mathrm{CuO_2}$ planes. The superconducting order parameter $\psi_m=\Delta_m \exp (i\phi_m)$ is defined on the sites $m$ of the square lattice where $\Delta_m$ and $\phi_m$ are the amplitude and phase respectively. The $\psi_m$ field is microscopically related to the complex spin-singlet pairing amplitude $\psi_m=\frac{1}{2}\langle a_{i\downarrow} a_{j\uparrow} -a_{j\downarrow} a_{i\uparrow}  \rangle$ on the $\mathrm{CuO_2}$ bonds where $m$ is the bond center of the nearest neighbour lattice sites $i$ and $j$ where $a_i(a^{\dagger}_i)$ are annihilation (creation) operators. The form of the functional $\mc{F}=\mc{F}_0+\mc{F}_1$
\begin{subequations}\label{Eq.functional}
\begin{eqnarray}
&&\mathcal{F}_0(\{\Delta_m\})=\sum_m \left(A\Delta_m^2 + \frac{B}{2}\Delta_m^4\right),\\
&&\mathcal{F}_1(\{\Delta_m,\phi_m\})=-C \sum_{\langle mn\rangle}  \Delta_m \Delta_n \cos(\phi_m-\phi_n-A_{mn}),~~~~~~
\end{eqnarray}
\end{subequations}

where $\langle mn\rangle$ denotes pairs of nearest neighbour bond sites and $A_{mn}(=\frac{2\pi}{\Phi_0}\int_m^n{\bf A}.d{\bf r}$) is the bond flux which incorporates the effect of an out of plane magnetic field. The motivation for these explicit forms of the parameters $A$, $B$ and $C$ from cuprate phenomenology and the details of temperature, doping dependence of a particular cuprate, e.g.~Bi2212  as discussed in \ref{app.Parameters}.
The form of the functional $\mathcal{F}\lbrace\phi_m,\Delta_m\rbrace$ is such that phase fluctuations are dominant and amplitude fluctuations weak at low doping $x$ and and become comparable in strength as $x$ increases ultimately tending towards Gaussian fluctuations of the full order parameter at large doping. The charge and heat current operators are (see \ref{app.Heat_charge_expressions})

\begin{eqnarray}
\rm{J}^{\rm e}=\frac{2\pi}{\Phi_0}C\Delta_m\Delta_n\sin(\phi_m-\phi_n-A_{mn})\\
\rm{J}^{\rm Q}=\frac{1}{2}(J^E_{m\rightarrow n}-J^E_{n\rightarrow m})+M_z(\bf{E}\times \hat{z})
\end{eqnarray}
where $\rm{J}^{\rm E}_{\rm {m}\rightarrow \rm{n}}=-\frac{C}{2}\lbrace \frac{\partial \psi^*_m}{\partial t} \sqrt{\frac{\psi_m}{\psi^*_m}}|\psi_n|e^{i\omega_{m,n}}+c.c.\rbrace$ with $\omega_{m,n}=\phi_m-\phi_n-A_{mn}$.

In the extreme type-II limit when the penetration depth $\lambda\rightarrow\infty$, the out of plane magnetic field $H$ is related to the in-plane bond flux $A_{mn}$ on a square plaquette $\Box$ of size $a_0$ such that $\sum_{\Box} A_{mn}=2\pi \frac{H a_0^2}{\Phi_0}$. The lattice constant $a_0$ introduces a field scale $H_0$ obtained when one flux quantum $\Phi_0$ passes through the square plaquette $\Box$ and $H_0=\frac{\Phi_0}{2\pi a_0^2}$. We also note that $\Delta_m\Delta_n\cos(\phi_m-\phi_n-A_{mn})=-(|\psi_m-\psi_n e^{iA_{mn}}|^2-\Delta_m^2-\Delta_n^2)$ and therefore the term $\mathcal{F}_1$ can be readily identified with the discretized version of the covariant derivative $|{\bf D}\Psi|^2$  in a standard Ginzburg-Landau theory. Thus, the lattice constant $a_0$ can be thought of as a suitable ultraviolet cutoff to describe the physics of the system.
  
\section{Simulation Geometry and Methodology}

\begin{figure}[htp]
\centering
\subfigure{\includegraphics[scale=0.3]{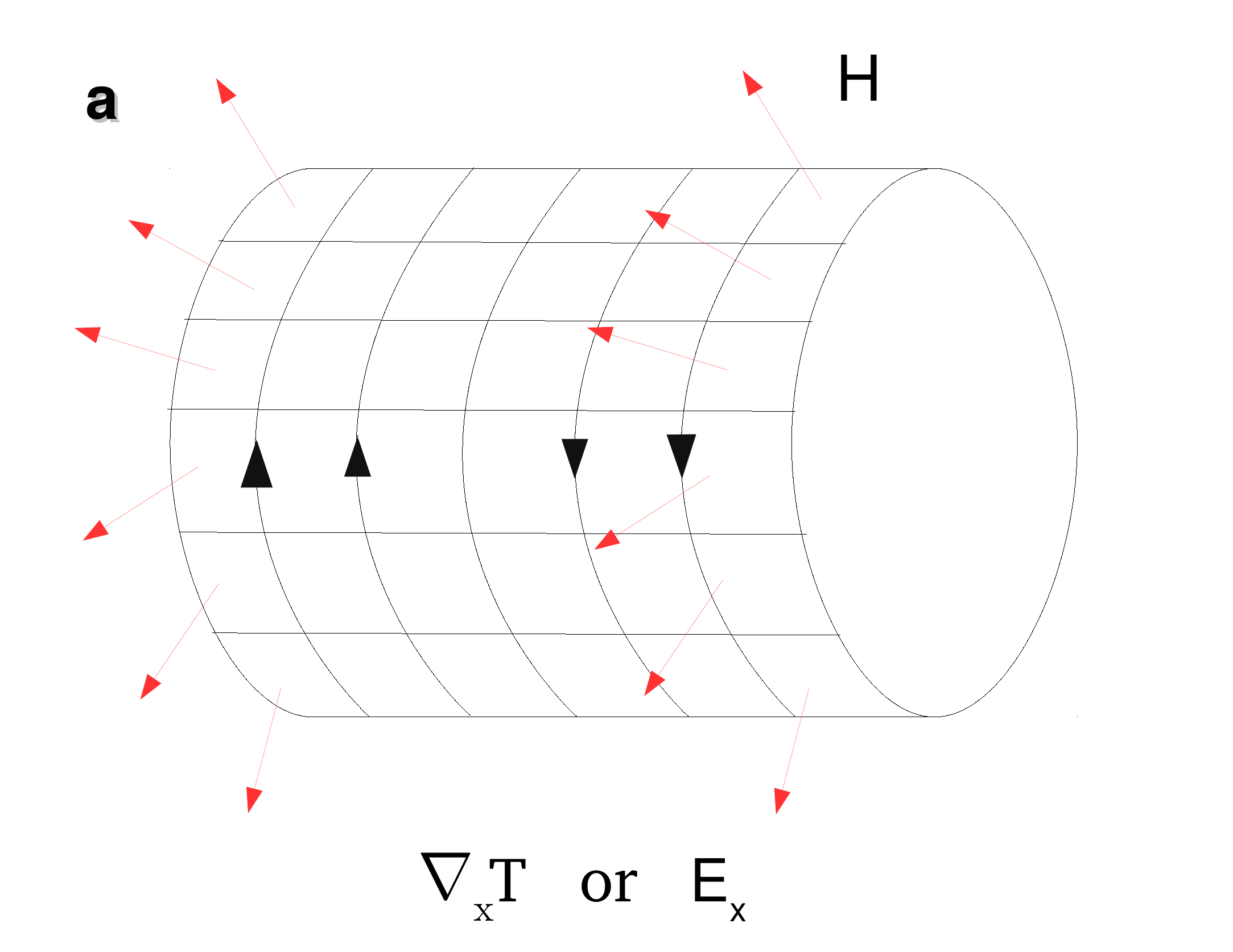}\label{fig:cylinder}}
\hspace{1cm}
\subfigure{\includegraphics[scale=0.3]{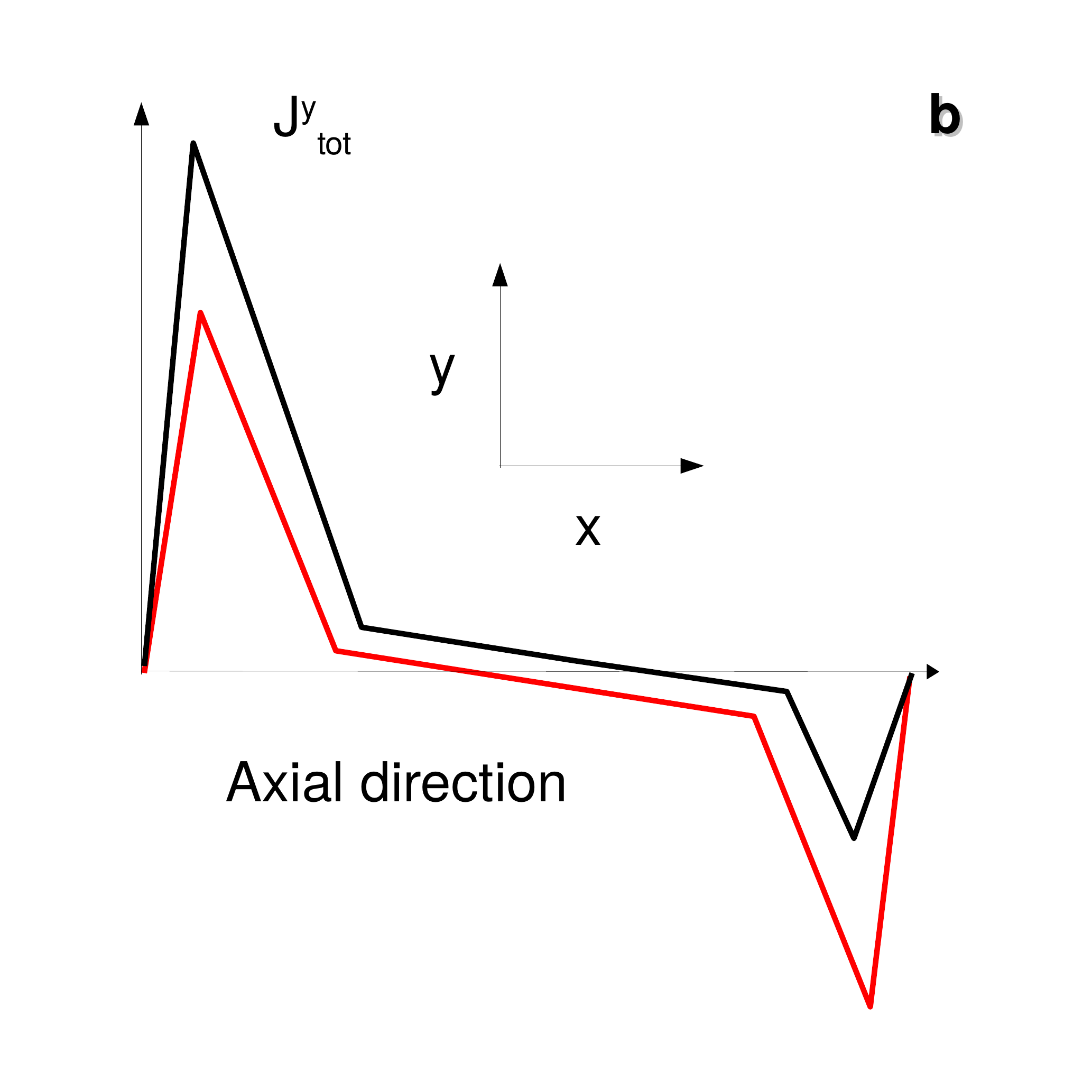}\label{fig:current}}
\caption{ a) The cylindrical geometry of our simulation. The magnetic field (${\bf H}$) is applied in the radially outward direction (red). The temperature gradient $\nabla_x T$ or $E_x$ is applied in the axial direction and the resulting current is in the azimuthal direction. b) The current profile of the cylinder in the presence and absence of a temperature gradient and electric field, which is shown by two different color (black and red) lines. The current density is maximum at the two edges of the cylinder. In the absence of a temperature gradient the current density is equal in magnitude at both edges of the cylinder (red line). When a temperature gradient is applied the current density increases at one end and decreases by the same amount at the other end (black line) }
\label{fig:Geometry_Current}
\end{figure} 


We simulate the model given by Eqn.~\ref{Eq.TDGL} numerically on a two dimensional system of size $100 \times 100$. We perform the simulation in dimensionless terms by scaling the relevant quantities by the units described in subsection 2.2. To compute $\alpha_{xy}$ we perform our simulations on a cylinder (Fig.~\ref{fig:cylinder}) with periodic boundary conditions in one direction ($\hat{y}$) and zero current conditions along the other ($\hat{x}$). The uniform magnetic flux per plaquette is in the radial direction and determined by the condition of zero flux in the axial direction. The resulting current is in the azimuthal direction and in the absence of any perturbations (temperature gradient, electric field etc) is maximum at the edges and falls to zero and changes direction at the center Fig.~\ref{fig:current} (red line). Thus, in the absence of any perturbing fields the background magnetization of the cylinder should be zero which can be checked by summing over the charge currents from one end to the other. 

A perturbing field like the temperature gradient along the axial direction introduces a transport current in the azimuthal direction and as a result the {\em {total}} current density is enhanced at one end and suppressed at the other (black line). We see this effect in our simulation by setting the temperature gradient in the linear response regime. Summing the total current density over the whole sample gives only the transport current since the sum over the magnetization current continues to be zero. $\alpha_{xy}$ can be obtained from the equation

\begin{equation}
\alpha_{xy}=-\frac{1}{S_A}\frac{\int J^e_{\rm tot} dS_A}{\nabla T}
\end{equation}
where $S_A$ is the area of the sample. The typical number of time steps chosen for equilibration and time averaging are about $1.2\times10^7$ and $10^6$ respectively.

We also compute the coefficient $\tilde{\alpha}_{xy}$ by switching off the temperature gradient and instead turning on the electric field ${\bf E}$ in the axial direction of the cylinder. ${\bf E}$ can be introduced through a time dependent magnetic vector potential (${\bf A}$) with ${\bf E}=-\frac{\partial{\bf A}}{\partial t}$, a position dependent electrostatic potential ${\bf E}=-\nabla \Phi$ or any gauge invariant combination of the two. In this method we calculate the total heat current density. It can be shown that the appropriate subtraction of the magnetization current to yield $\tilde{\alpha}_{xy}$ gives
\begin{equation}
\tilde{\alpha}_{xy}=-(\frac{1}{S_A}\frac{\int J^Q_{tot} dS_A}{E}-M)
\end{equation}
The magnetization ${\bf M}$ is obtained from ${\bf J}^{\rm {e}}_{\rm {mag}} = \boldnabla {\bf \times} {\bf M}$ by an appropriate integration in the equilibrium state (i.e. zero electric field and temperature gradient). The values obtained are in agreement with those from Monte-Carlo simulations obtained in a previous study~\cite{Sarkar_2016}.

A check for whether the magnetization current subtraction has been done properly is by verifying the equality $\alpha_{xy}=\frac{\tilde{\alpha}_{xy}}{T}$, which is a consequence of the Onsager relations for transport coefficients. We have verified that the above equality holds to within our noise levels for all values of doping, temperature and field. We note that the in the underdoped region, where the fluctuations are strong, there is a large separation between $T_c$ and $T_c^{MF}$. Thus, fluctuations of the amplitude of $\Psi$ are negligible even up to temperatures significantly greater than $T_c$ (but also significantly lower than $T_c^{MF}$). This allows us to use an effective $XY$ model with only a dynamically varying phase and amplitude frozen to the mean-field value up to fairly high temperatures at underdoping. This effective $XY$ model seems to have a lower noise level for $\alpha_{xy}$ as compared to the full Ginzburg-Landau model. We thus employ this effective model for lower noise in the underdoped region and have verified that the results agree with those obtained from the full model to within error bars.

\section{Results}

\begin{figure}[H]
\begin{tabular}{ccc}
\includegraphics[scale=0.20]{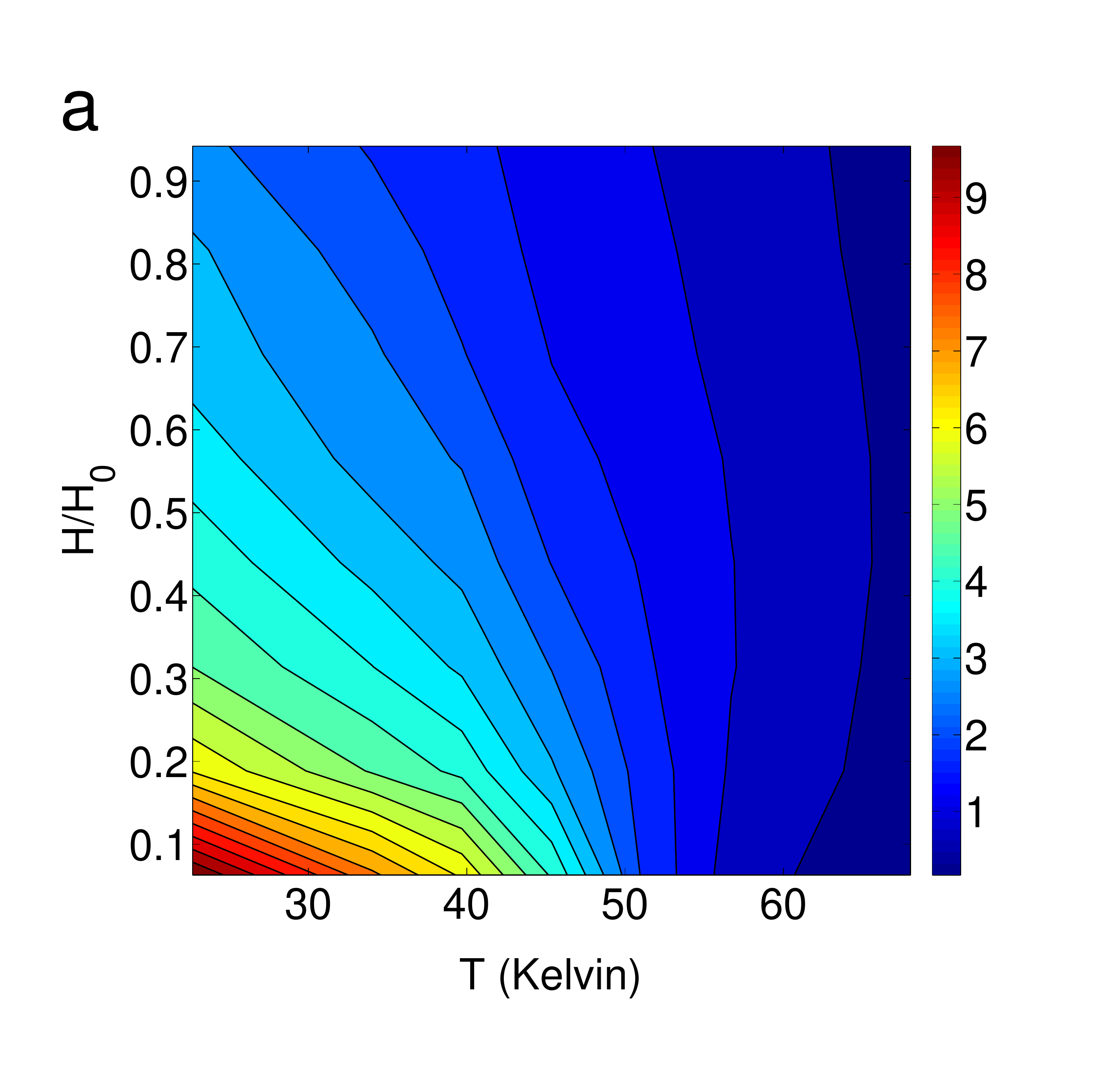}&\includegraphics[scale=0.20]{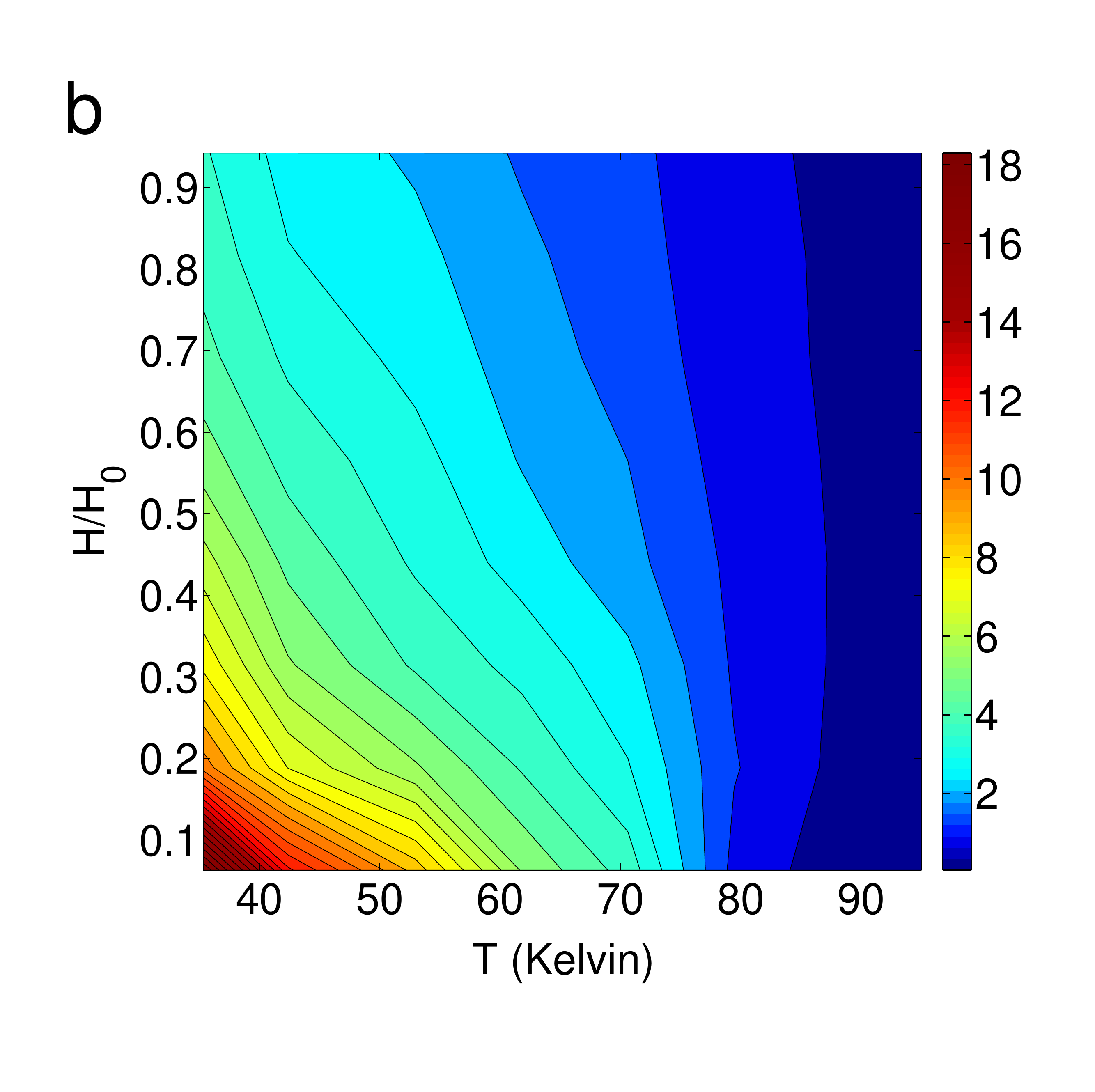}&\includegraphics[scale=0.20]{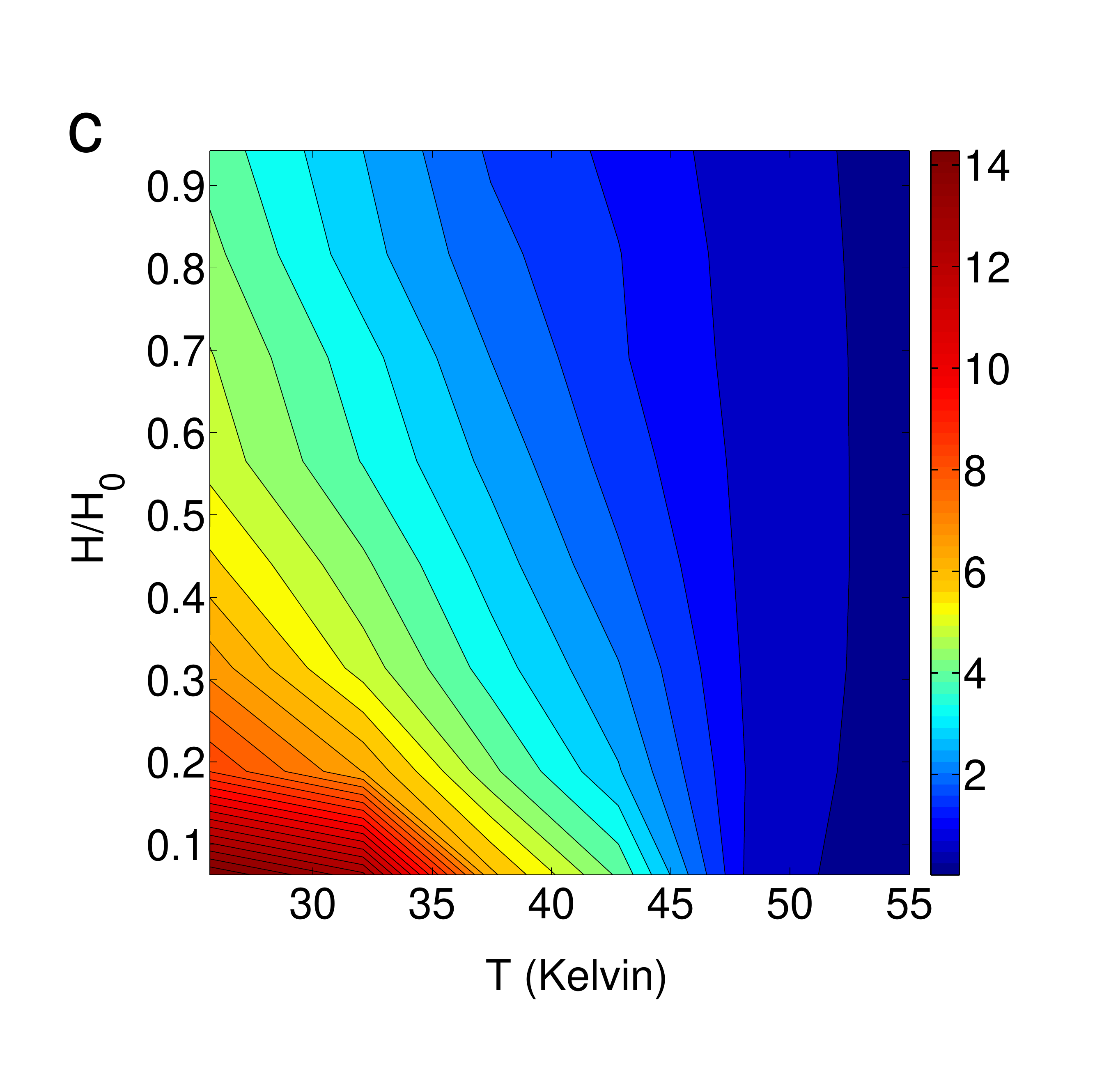}
\end{tabular}
\caption{Contour plots of  $\alpha_{xy}$ of a) UD(x=0.05), b) OPT(x=0.15) and c) OD(x=0.25) cuprates in SI units. The contour lines are almost vertical near $T\geq T_{c}$, analogous to previously obtained magnetization contour lines in Fig 2 of Ref.~\cite{Sarkar_2016} and also consistent with the features obtained by Podolsky et al~\cite{Podolsky}.}
\label{fig:Contour}
\end{figure}

We plot the obtained values of $\alpha_{xy}$ as functions of doping, temperature and field. The overall features of $\alpha_{xy}$ over the phase diagram are summarized in Fig.~\ref{fig:Contour} through color map plots of the strength of the $\alpha_{xy}$ in the field-temperature ($H-T$) plane for three different values of doping going from underdoped to overdoped. We have also compared $\alpha_{xy}$ to ${\bf M}$. ${\bf M}$ can in turn be compared directly to experiments as was done by us in a previous study based on the model we employ here~\cite{Sarkar_2016}. We found the calculated {\bf M} to be in reasonably good quantitative agreement across the entire range of doping, field and temperature accessible in experiments on the cuprates~\cite{Ong_2006,Li_2010}.  The value of $\alpha_{xy}$ for our two dimensional system is converted to a three dimensional one by dividing by the lattice spacing of BSSCO to enable a direct comparison to the three dimensional magnetization.

\begin{figure}[H]
\begin{tabular}{cc}
\includegraphics[scale=0.23]{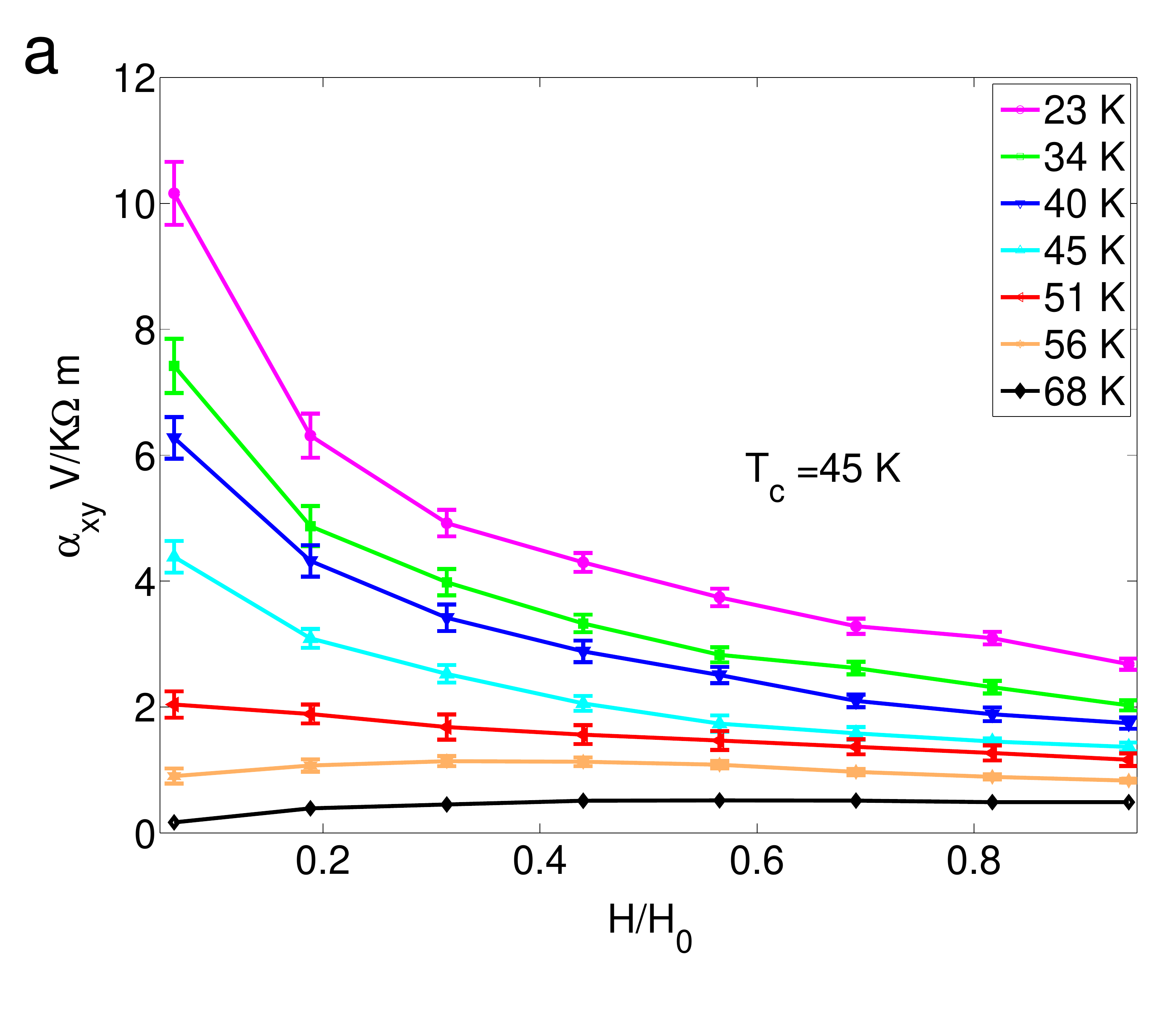}&\includegraphics[scale=0.23]{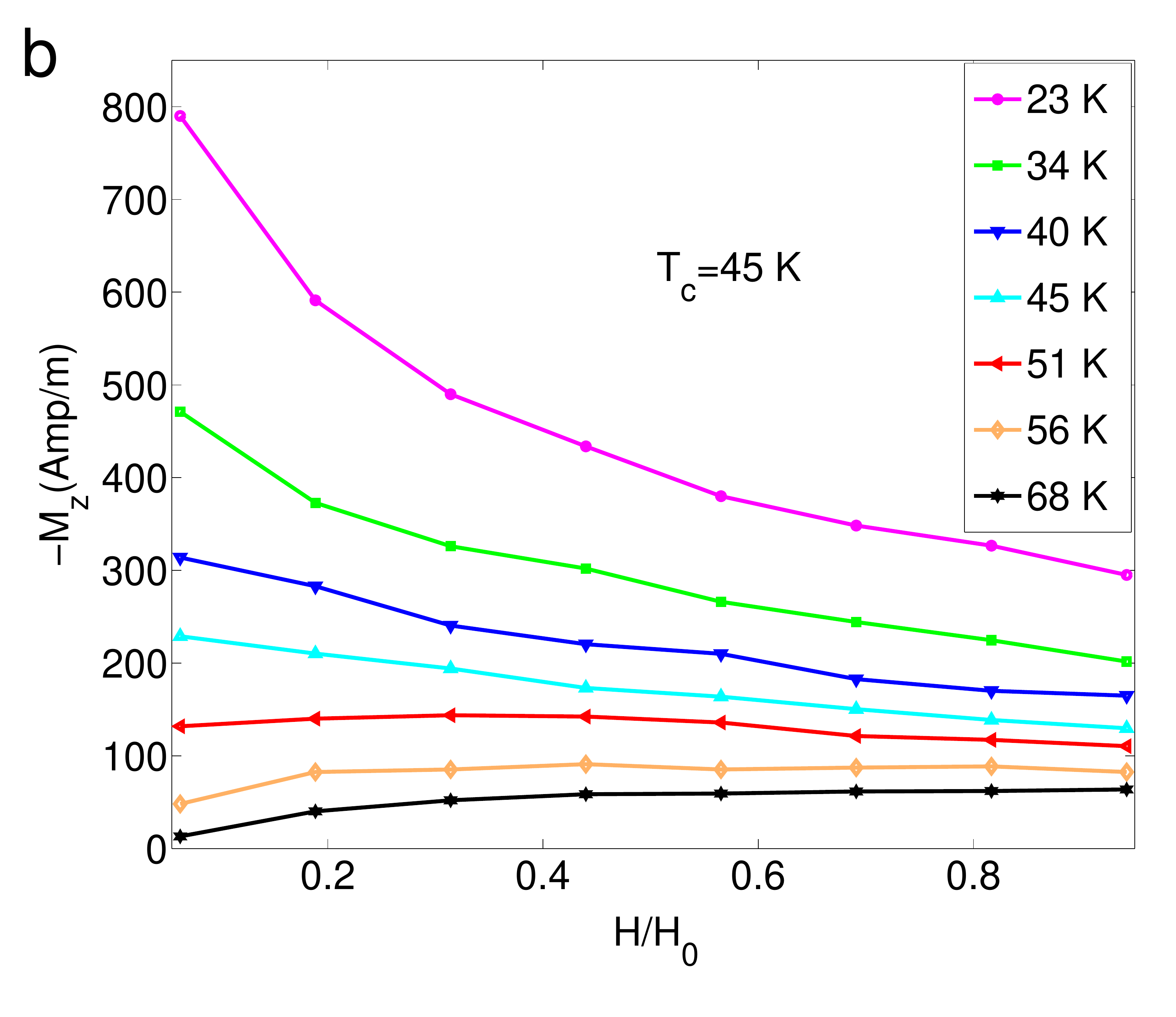}
\end{tabular}
\begin{tabular}{cc}
\includegraphics[scale=0.23]{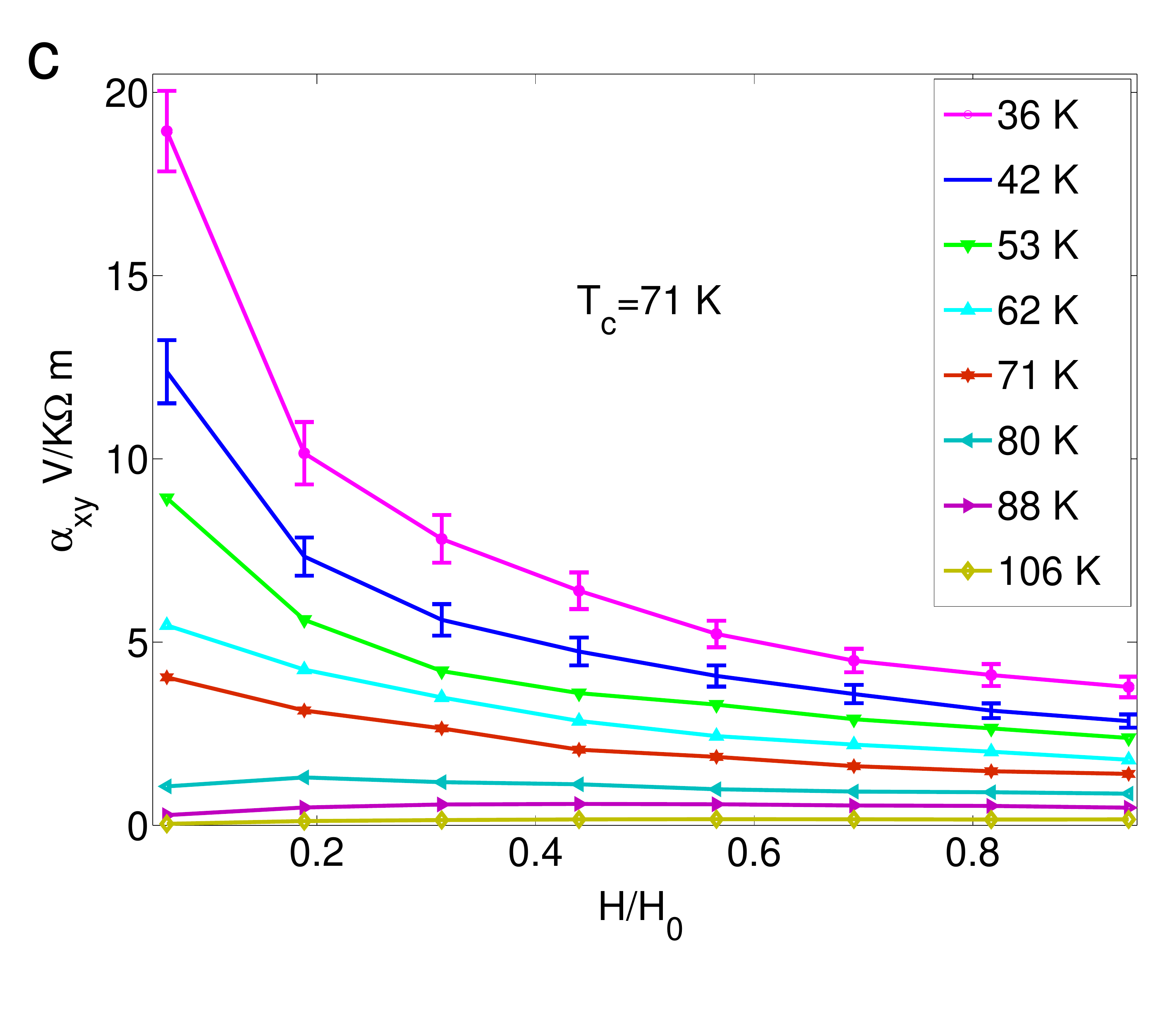}&\includegraphics[scale=0.23]{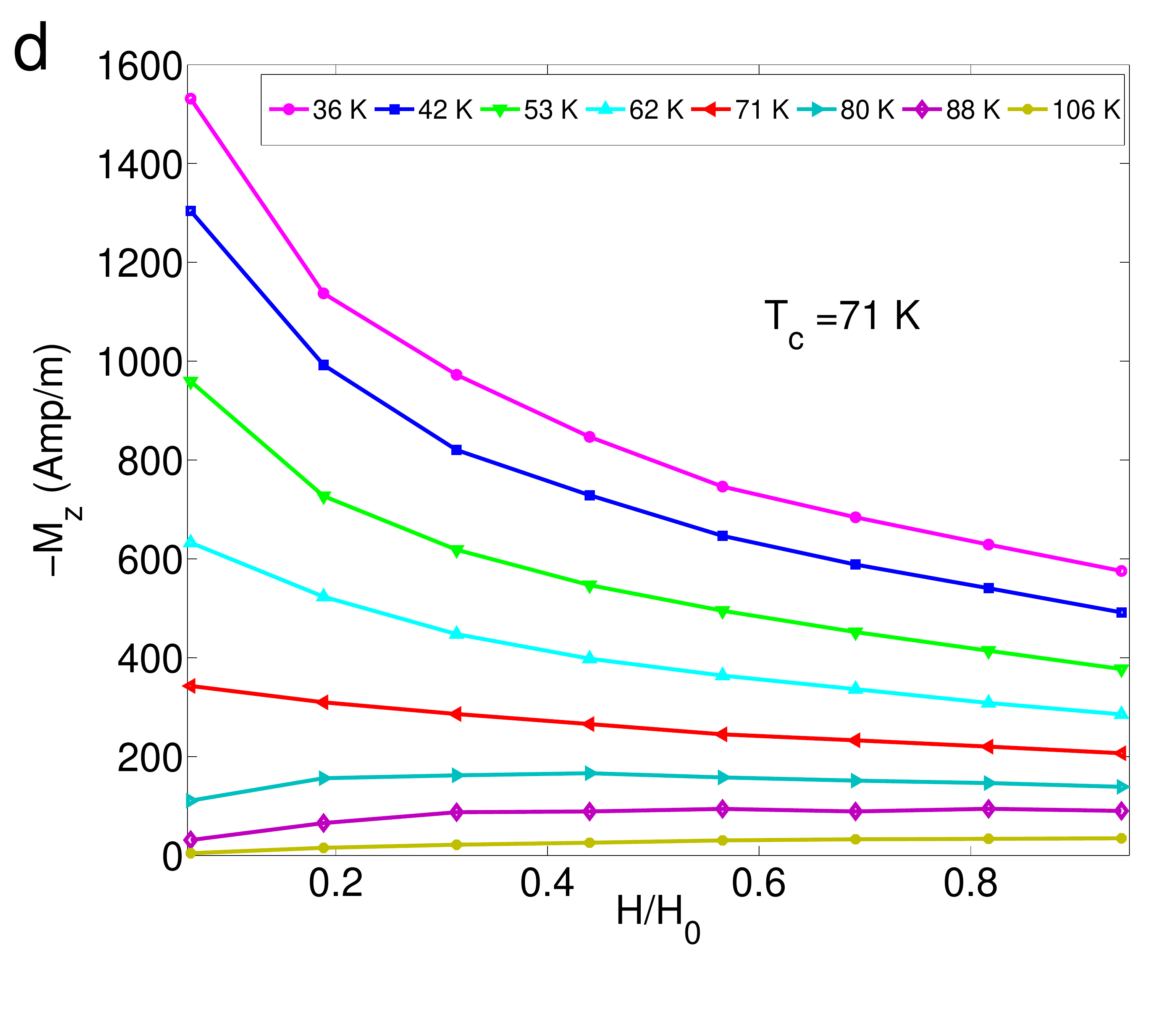}
\end{tabular}
\begin{tabular}{cc}
\includegraphics[scale=0.23]{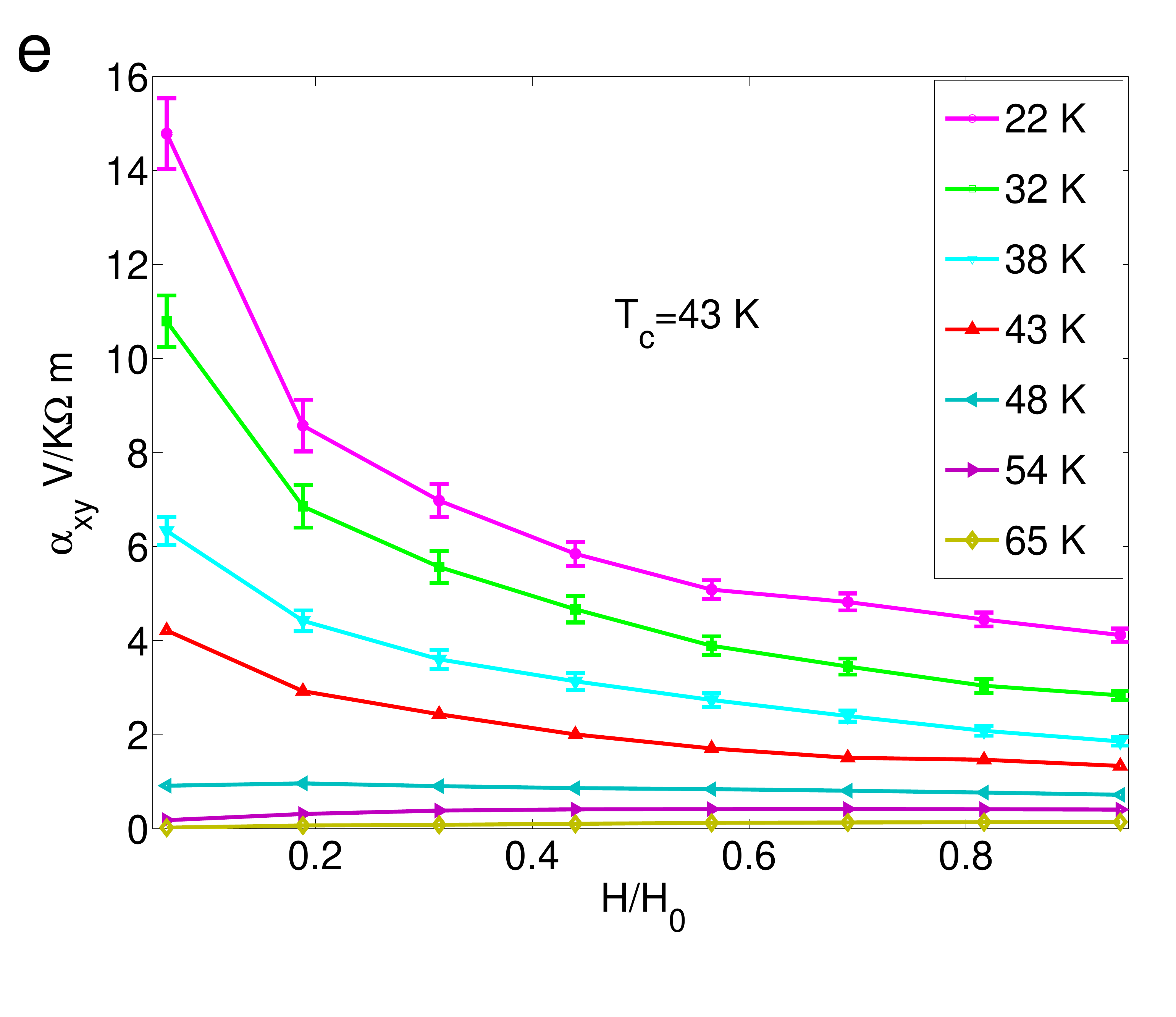}&\includegraphics[scale=0.23]{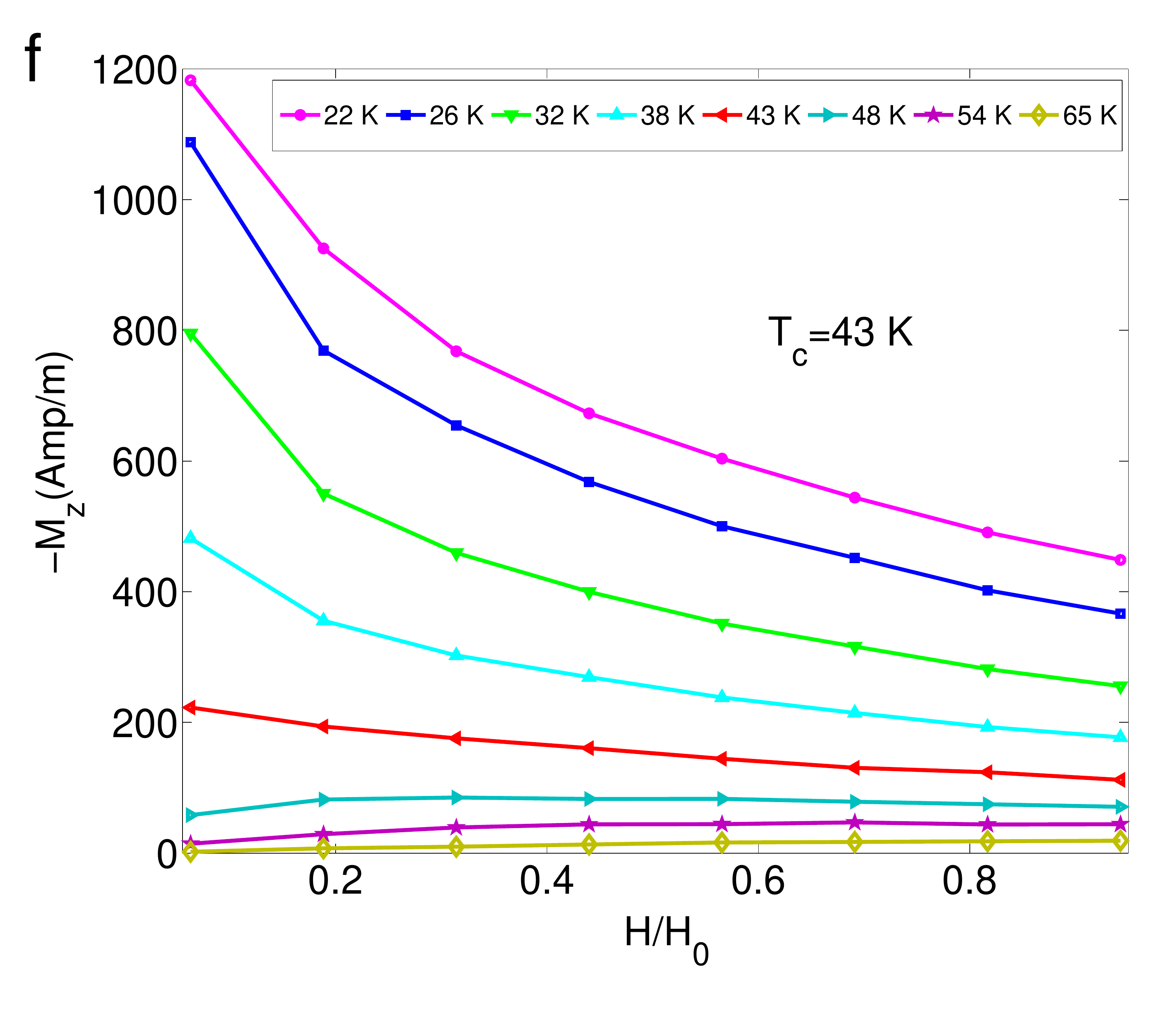}
\end{tabular}
\caption{The field dependence (in units of $\mathrm{H_0}$) of $\alpha_{xy}$ and magnetization $-M$  for a) and b) underdoped (UD) (x=0.05), c) and d) optimally doped (OPT) (x=0.15), e) and f) overdoped (OD) (x=0.25) cuprates in SI units ($\rm{V/K\Omega m}$ and Amp/m respectively). We divide the numerically obtained $\alpha_{xy}^{2d}$ and $M^{2D}$ by appropriate layer spacing  $ d=1.5~\mathrm{nm}$ to convert to the three dimensional $\alpha_{xy}$ and $M$. $\alpha_{xy}$ and $M$ can be seen to behave in the same way as a function of field at different temperatures. As $H\to 0$, both $\alpha _{xy}$ and $M$ diverge for $T<T_c$ and go to zero for $T>T_c$.}
\label{fig:Field}
\end{figure}

Fig.~\ref{fig:Field} shows the field dependence of $\alpha_{xy}$ at different temperatures for three representative values of doping - one each in the underdoped, optimally doped and overdoped regimes, with respective $T_c$ values indicated in the figure panels. The magnetization ${\bf M}$ is shown alongside to enable a comparison. It can be seen that the overall dependence on temperature and field is the same for both quantities for all three values of doping. This is significant because the strength of superconducting fluctuations is different for the three regimes going from strong to weak as the value of doping increases. This similarity of the gross features in the field and temperature dependence of both quantities is a consequence of the fact that it is the strength of the superconducting fluctuations rather than their dynamics that is responsible for both the diamagnetic and off-diagonal thermoelectric responses. The color plots of $\alpha_{xy}$ in Fig.~\ref{fig:Contour} illustrate the field and temperature dependence better making it possible to identify contours of constant $\alpha_{xy}$.


\begin{figure}
\begin{tabular}{cc}
\includegraphics[scale=0.24]{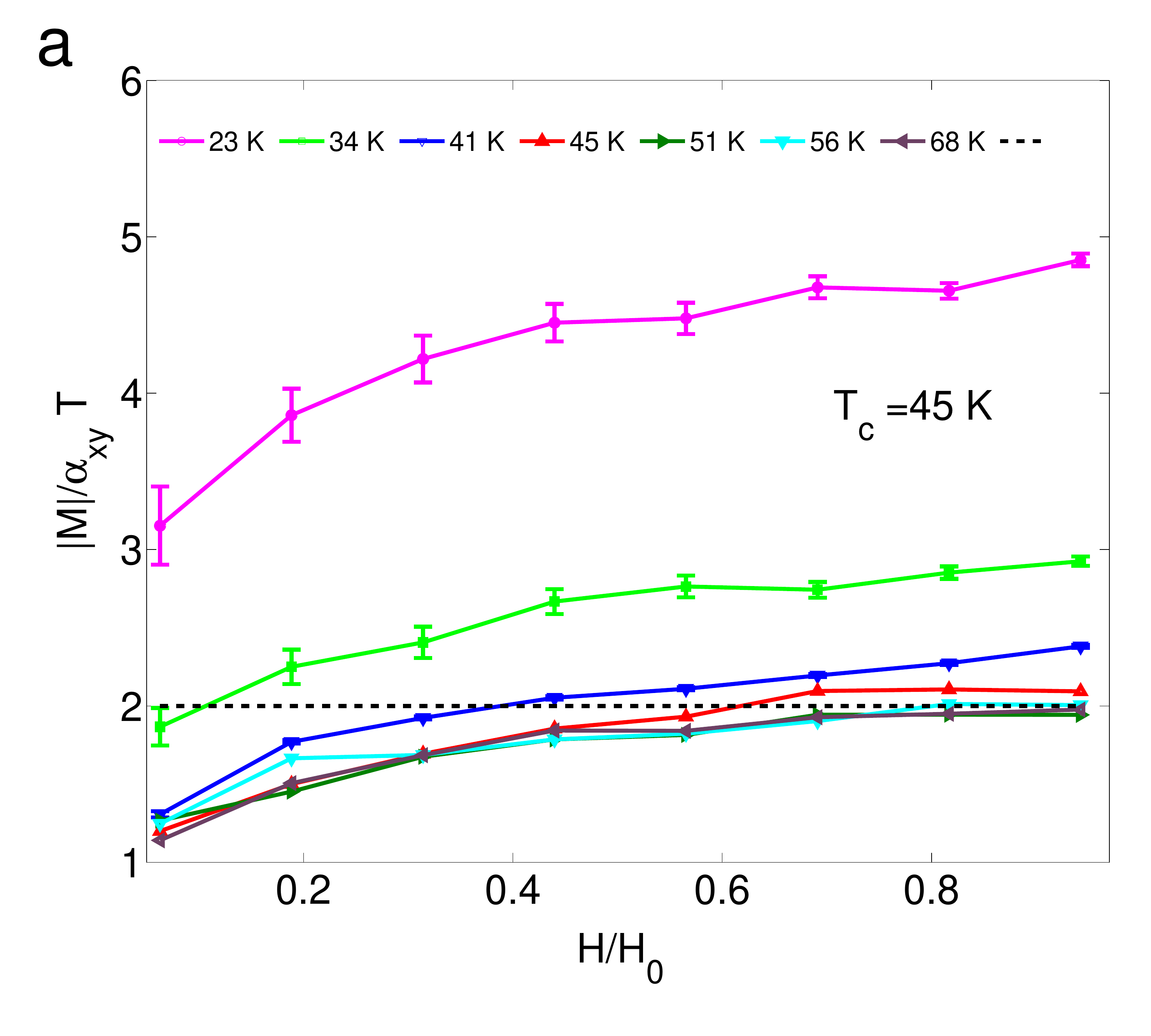}&\includegraphics[scale=0.24]{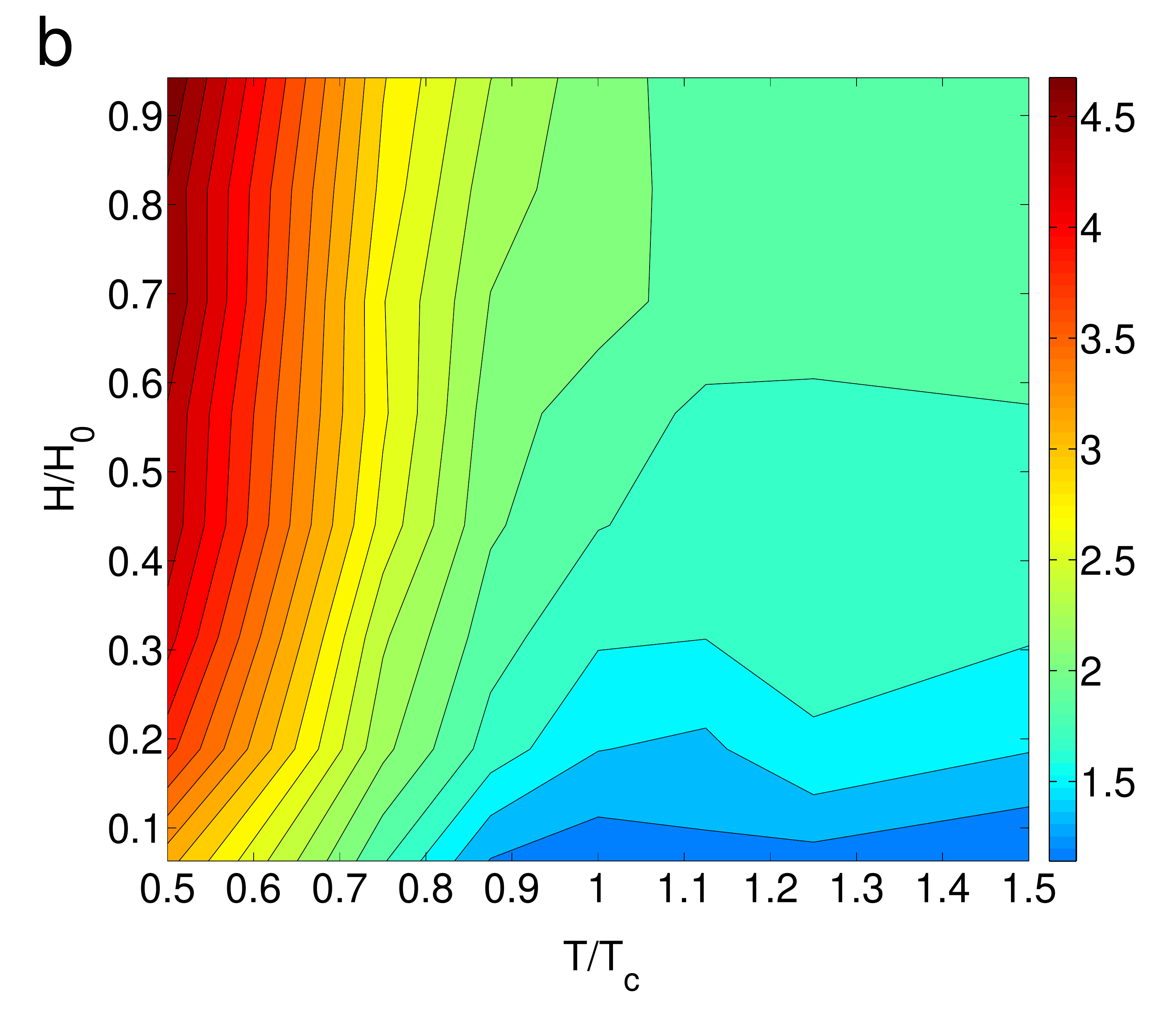}\\
\includegraphics[scale=0.24]{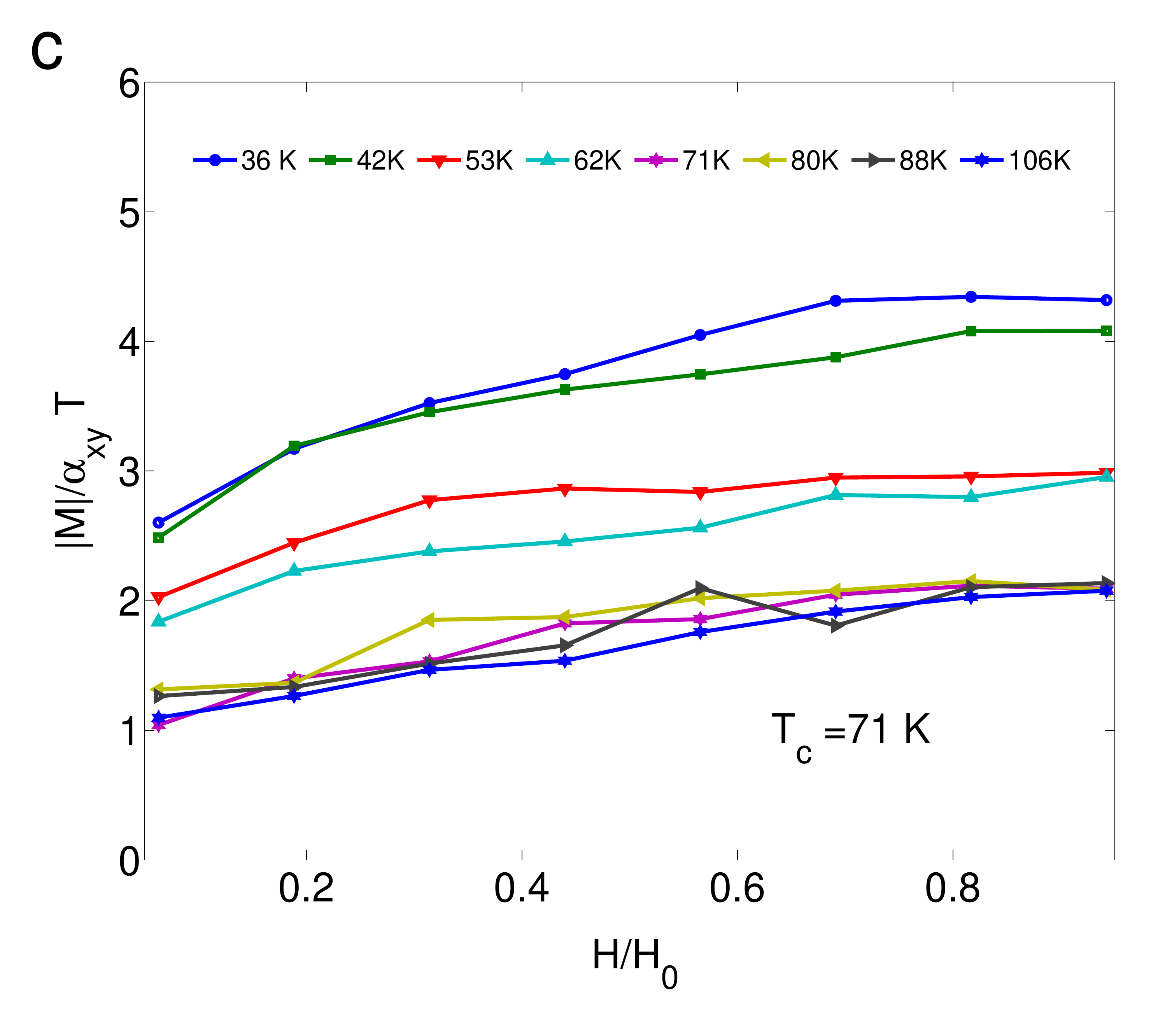}&\includegraphics[scale=0.24]{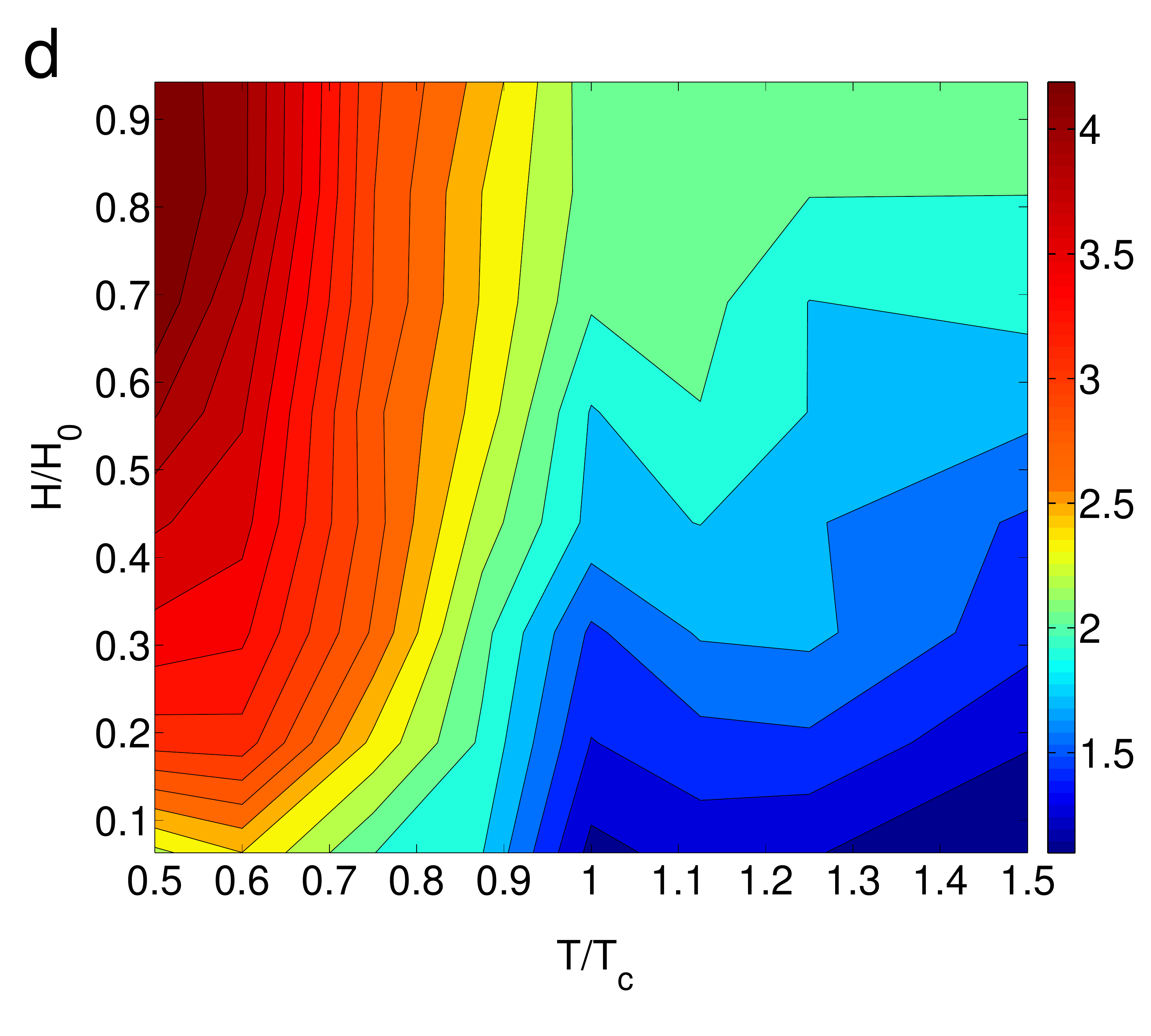}\\
\includegraphics[scale=0.24]{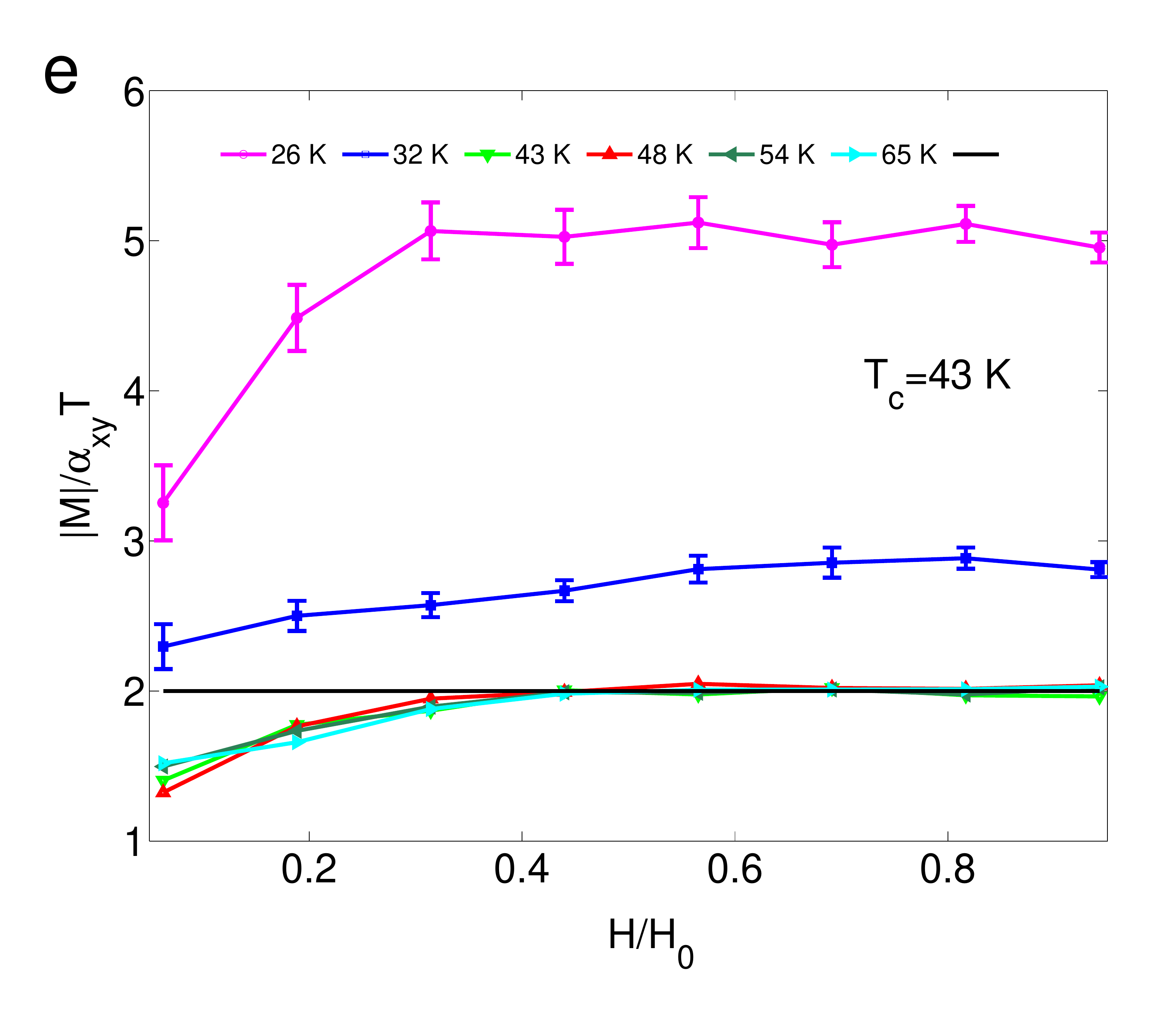}&\includegraphics[scale=0.24]{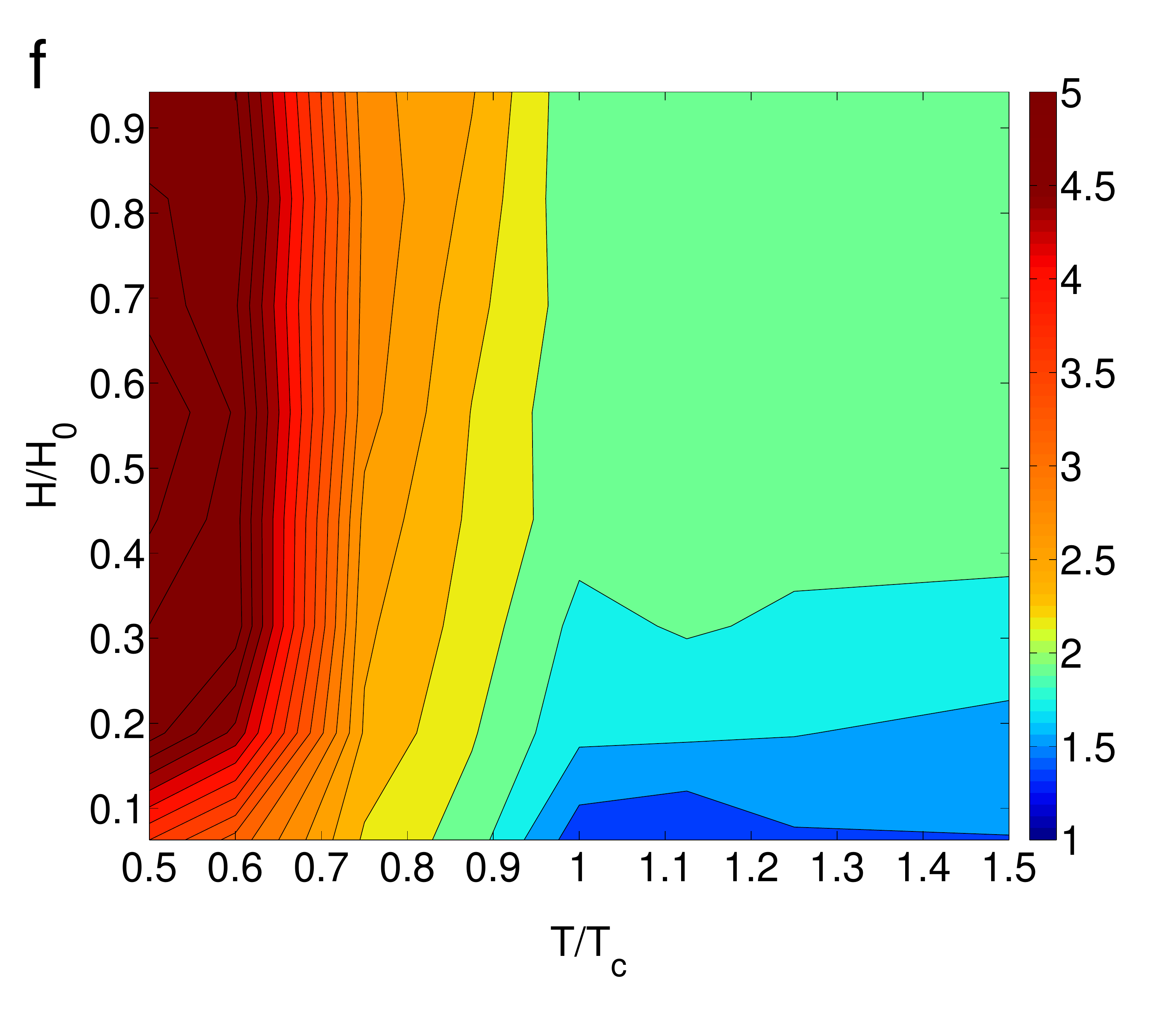}
\end{tabular}
\caption{The dimensionless quantity $|{\bf M}|/\alpha_{xy}T$ is obtained at different temperatures for a) UD ($x=0.05$), c) OPT ($x=0.15$) and e) OD ($x=0.25$) cuprates. The data shows at high temperature $|{\bf M}|/\alpha_{xy}T\simeq 2$. Colormap contour lines of the dimensionless quantity $|{\bf M}|/\alpha_{xy}T$ in the $H-T$ plane for b) UD  ($x=0.05$) d) OPT ($x=0.15$) and e) OD ($x=0.25$) region. The temperature axis is scaled in units of $T_c$. The dimensionless quantity $\frac{|{\bf M}|}{\alpha_{xy}T}$ behaves similarly at high temperature (compared to $T_c$) for different values of doping ranging from underdoped to overdoped.}
\label{fig:fig}
\end{figure}

The similarity between the field and temperature dependences of $\alpha_{xy}$ and ${\bf M}$ motivates a more careful comparison of the two quantities. As argued in the previous section, the quantity $|{\bf M}|/(T \alpha_{xy})$ is dimensionless and hence a good measure of the correlations between the two quantities ${\bf M}$ and $\alpha_{xy}$.  Plots of this quantity are shown in Fig.~\ref{fig:fig} and it can be seen that it is not a constant but has a dependence on doping $x$, temperature $T/T_c$ and field $H/H_0$. Of particular relevance is the fact that it stays close to the value $2$ for $T>T_c$ at both underdoping and overdoping over a substantial range of field as shown in Figs.\ref{fig:fig}(a),(e). This is consistent with the predictions of theoretical calculations in the high temperature limit of the $XY$ model and the Guassian fluctuation limit respectively as we discuss in the next section~\cite{Podolsky, Ussishkin}. The dimensionless ratio has also been calculated to be $2$ for a model with both superconducting and charge density wave order~\cite{Orgad2_2014}. For optimal doping, the ratio approaches 2 at high fields in our numerical calculations. It should be noted that the ratio appears to be less than $2$ at low fields. This is consistent with results obtained from self-consistent Gaussian fluctuations\cite{Tinh_2014}. However, the signal to noise ratio in the simulations at low fields is small and we can not infer anything conclusively about the ratio $|{\bf M}|/(T \alpha_{xy})$ in this regime.

\begin{figure}
\centering
\includegraphics[height=6cm]{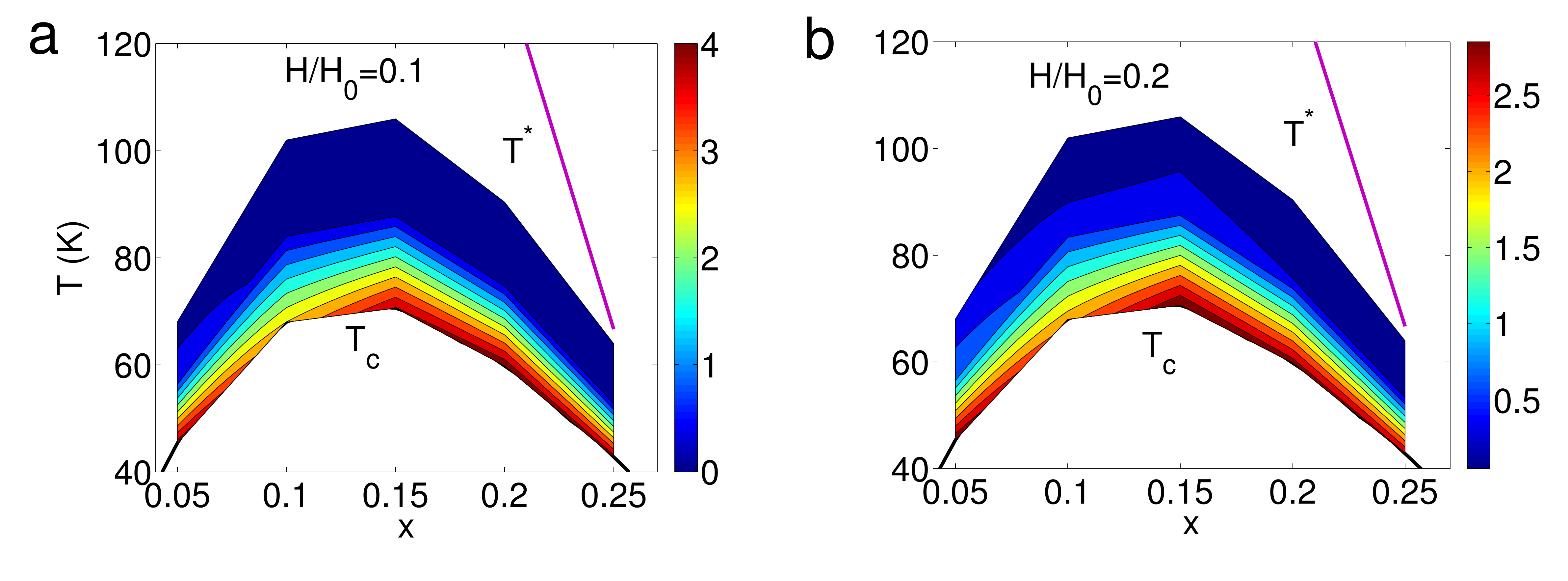}
\caption{$\alpha_{xy}$ in the $x-T$ plane for two different values of the magnetic field (a) $H/H_0=0.1$ and (b) $H/H_0=0.2$. The lines of constant $\alpha_{xy}$ follow the superconducting dome. This indicates that the equilibrium superconducting fluctuations responsible for the suppression of the superfluid stiffness also determine the thermoelectric response. A similar feature is also seen for the magnetization~\cite{Sarkar_2016}.}\label{fig:Nern_PD}
\end{figure}

A final feature of our simulation data that needs to be highlighted is shown in Fig.~\ref{fig:Nern_PD}. In this figure contours of constant $\alpha_{xy}$ are plotted in the $x-T$ plane for different values of the magnetic field for $T>T_c$. The superconducting dome obtained by calculating $T_c$ as a function of $x$ is also plotted. It can be seen that the contours follow the superconducting dome. This is especially significant at underdoping where the transition temperature is determined by the strength of phase fluctuations that in turn suppress the superfluid stiffness. We discuss the relevance of this feature in our data in the next section, but note that the same feature is also seen in the fluctuation diamagnetism experimentally~\cite{Li_2010,Xiao} and in theoretical calculations~\cite{Sarkar_2016}. More significantly, the same feature has also been seen in experimental data for the Nernst coefficient~\cite{Ong_2006}.

\section{Discussion and conclusions}
 
 We have obtained $\alpha_{xy}$ and the magnetization ${\bf M}$ as functions of temperature and magnetic field from a phenomenological model of superconducting fluctuations. This model is described by a Ginzburg-Landau free energy on a lattice with the coefficients of the different terms chosen as parameters of the temperature and doping to reproduce several experimentally observed equilibrium properties of the cuprates. Transport is modeled by introducing simple relaxation dynamics for the superconducting order parameter. Correlations between the Nernst signal and the diamagnetism have been observed in experiments. The Nernst signal is $\alpha_{xy}/\sigma_{xx}$ for systems with small values of the Hall angle and thermopower, as is the case for the cuprates over large parts of the phase diagram. We have argued here that the correlation between the Nernst signal and the magnetization arises primarily due to a correlation between $\alpha_{xy}$ and the magnetization in a model with only superconducting fluctuations since both quantities depend only on the strength of the fluctuations and not their dynamics. The relationship between $\alpha_{xy}$ and  {\bf M} is quantified by calculating the the dimensionless ratio ${\bf M}/(T \alpha_{xy})$. This ratio has been calculated by other authors previously for a model of superconducting fluctuations in the $XY$ limit of strong phase fluctuations and the Gaussian limit and found to be equal to $2$ in both~\cite{Ussishkin, Podolsky}. These correspond to high temperature limits $T \gg T_c$ for the overdoped and underdoped cuprates respectively. Here, we have calculated this ratio as a function of field, temperature and doping for the entire phase diagram and found deviations from the value of $2$ in regions where the high temperature approximation does not apply. 
 
$\alpha_{xy}$ calculated as a function of temperature, field and doping is shown is Figs.~\ref{fig:Field},\ref{fig:Contour} alongside ${\bf M}$. It can be seen that the dependence of both quantities on field and temperature is very similar for the entire range of doping. This has previously been demonstrated in certain limits for very underdoped and overdoped samples~\cite{Ussishkin,Mukerjee,Podolsky}. Our calculations agree with these previous results. On the underdoped side, our model reduces to a phase only model for a large range of temperatures for which the amplitude of the superconducting order parameter is effectively constant with no spatial or temporal fluctuations. This corresponds to the $XY$ limit which was the subject of one of the aforementioned studies~\cite{Podolsky}. On the overdoped side, the strength of the fluctuations is weaker resulting in a smaller difference between $T_c$ and $T_c^{MF}$. In this limit both phase and amplitude fluctuate together and cannot be disentangled from each other. The description of the physics of the system is thus in terms of fluctuations of the full order parameter. At high temperature, the system is in the Gaussian limit and our results agree with previous calculations of $\alpha_{xy}$ in low fields in this limit~\cite{Ussishkin}. At higher fields too in the overdoped limit, our calculations agree with previous work~\cite{Mukerjee}. 
 
 One of the new results of our work is that we have shown that one can smoothly interpolate between these previously studied limits by employing the free energy functional~\eqref{Eq.functional} to calculate $\alpha_{xy}$. As a result, we are able to directly show the connection not just between $\alpha_{xy}$ and ${\bf M}$ but also between these quantities and others whose nature is primarily determined by superconducting fluctuations, across the entire phase diagram. One of these quantities is the superfluid stiffness, the disappearance of which corresponds to the destruction of superconductivity at the transition temperature $T_c$. The correlation between $\alpha_{xy}$ and $T_c$ can be seen in Fig.~\ref{fig:Nern_PD} where curves of constant $\alpha_{xy}$ in the temperature and doping plane follow the superconducting dome for different values of the magnetic field. A similar correlation also exists between ${\bf M}$ and $T_c$, which we have shown in an earlier work~\cite{Sarkar_2016}.  
 
 The ratio $\mathbf{M}/(T \alpha_{xy})$ is plotted in Fig.~\ref{fig:fig} for different values of temperature, field and doping. It has been remarked earlier that this value has been shown to be equal to $2$ at high temperature for the $XY$ model~\cite{Podolsky} and in the limit of Gaussian fluctuations at low field~\cite{Ussishkin}. Our model extrapolates to both limits for appropriate choices of parameters but we have to be careful in defining what we mean by high temperature. The $XY$ limit is obtained when the separation between $T_c^{MF}$ and $T_c$ becomes large, which corresponds to underdoping. High temperature here means temperatures large compared to $T_c$ but small compared to $T_c^{MF}$. This defines a fairly wide range of temperatures since the two scales are well separated. On the other hand, the Gaussian limit corresponds to a small separation between $T_c$ and $T_c^{MF}$ (overdoping) and high temperature here means a temperatures large compared to both. It should be emphasized that there is a Gaussian regime for any value of doping for temperatures larger than $T_c^{MF}$. However, for underdoped systems, these temperatures are much higher than the ones at which experimental measurements are performed and are thus not relevant here. Optimally doped systems lie in neither regime and our work provides the first calculation of the ratio ${\bf M}/(T \alpha_{xy})$ for them. Even in the underdoped and overdoped regime, we calculate for the first time the ratio beyond the high temperature limits discussed above. It can be seen that ${\bf M}/(T \alpha_{xy})$ agrees with the previously obtained results mentioned above. 
 
 It is interesting to note that while ${\bf M}/(T \alpha_{xy})$ obtained from our simulations does deviate from the value of $2$ at low temperatures (See Fig.~\ref{fig:fig}), it attains this ``high temperature'' value even at temperatures comparable to $T_c$. In fact for the underdoped system, it does so even at temperatures lower than $T_c$. Thus, it appears that in so far as this quantity is concerned, the Gaussian regime ($T \gg T_c^{MF}$) is not distinguishable from the strongly phase fluctuating regime. We emphasize that this does not imply that the two regimes are indistinguishable for each of the two quantities ${\bf M}$ and $\alpha_{xy}$ individually. Indeed, the temperature dependence of the these two quantities at low field has been shown to be distinct in the two regimes~\cite{Podolsky, Ussishkin} but their ratio appears to not make that distinction since the leading temperature dependence cancels between the numerator and the denominator. Thus, there does not seem to be a very clear distinction between the underdoped, optimally doped and overdoped systems with the temperature scale for the ratio being set only by $T_c$ regardless of whether $T_c^{MF}$ is in its vicinity. We note that the value of ${\bf M}/(T \alpha_{xy})$ appears to be less than $2$ at high temperature for the lowest fields.
This could be an artifact of high noise levels in this regime and a higher precision calculation (which would be fairly time consuming) may yield a value equal to $2$. 
 
 The utility of our calculation is in identifying the correlation between the magnetization ${\bf M}$ and $\alpha_{xy}$. For a superconducting system, a strong diamagnetic signal, even above $T_c$ is typically due to superconducting fluctuations as opposed to other excitations like quasiparticles~\cite{Ghosal_2007}. However, the Nernst signal, can have substantial contributions from these other excitations in addition to from superconducting fluctuations. In fact, the role of quasiparticles in the observed large Nernst effect of the cuprates has been discussed extensively in Refs.\cite{Taillefer_2010,Sachdev_2010}. Our calculation provides a method for determining the extent of the contribution of superconducting fluctuations to the observed Nernst signal through the ratio of ${\bf M}/(T \alpha_{xy})$. If the observed ratio is close to the predictions from our model then superconducting fluctuations are chiefly responsible for the Nernst effect in the particular regime of temperature, field and doping. We would likely to emphasize again that the relevant transport quantity in our calculation is $\alpha_{xy}$ and not the Nernst signal $\nu$. Experimentally, obtaining $\alpha_{xy}$ requires a concurrent measurement of the Nernst effect and the magnetoconductance. It is also possible that features in the Nernst effect unconnected to superconducting fluctuations, and hence the magnetization, arise due to the behavior of the magnetoconductance and not $\alpha_{xy}$. An analysis of these features is beyond the scope of a calculation like ours.
 
 To summarize, we have studied the Nernst effect in fluctuating superconductors by calculating the transport coefficient $\alpha_{xy}$. We have employed a phenomenological model of superconducting fluctuations in the cuprates, which allows us to calculate $\alpha_{xy}$ and the magnetization ${\bf M}$ over the entire range of experimentally accessible values of field, temperature and doping. We have found fairly good agreement with experimental data, wherever available and previous theoretical calculations in specific regimes of the parameters. We have argued that $\alpha_{xy}$ and ${\bf M}$ are both determined by the equilibrium properties of the superconducting fluctuations (and not their dynamics) despite the former being a transport quantity. Consequently, there exists a dimensionless ratio ${\bf M}/(T \alpha_{xy})$ that quantifies the relation between the two quantities. We have calculated this ratio over the entire phase diagram of the cuprates and found that it agrees with previously obtained results. Further, it appears that there is no sharp distinction between phase fluctuations and Gaussian fluctuations for this ratio even though there is for $\alpha_{xy}$ and ${\bf M}$ individually. The utility of this ratio is that it can be used to determine the extent to which superconducting fluctuations contribute to the Nernst effect in different parts of the phase diagram given the measured values of magnetization.
 
 \section{Acknowledgements} 
K.S. would like to thank CSIR (Govt. of India) and S.M. thanks the DST (Govt. of India) for support. T.V.R. acknowledges the support of the DST Year of Science Professorship, and the hospitality of the NCBS, Bangalore. The authors would like to thank Subhro Bhattacharjee for many stimulating comments and discussions.

\appendix
\section{The free energy functional} \label{app.Parameters}
The functional form in the absence of a gauge field is defined as 
\begin{subequations}\label{Eq.functional2}
\begin{eqnarray}
&&\mathcal{F}_0(\{\Delta_m\})=\sum_m \left(A\Delta_m^2 + \frac{B}{2}\Delta_m^4\right),\\
&&\mathcal{F}_1(\{\Delta_m,\phi_m\})=-C \sum_{\langle mn\rangle}  \Delta_m \Delta_n \cos(\phi_m-\phi_n),~~~~~~
\end{eqnarray}
\end{subequations}
where the pairing field $\psi_m=\Delta_m \exp(i\phi_m)$ is defined on the sites $m$ of the square lattice with phase $\phi_m$ and amplitude $\Delta_m$. $\langle mn\rangle$ denotes nearest neighbour site pairs. 
The coeffecients $A$, $B$ and $C$ are given doping $x$ and temperature $T$ dependence from cuprate experiments in a phenomenological way with dimensionless numbers $f$, $b$, $c$ and a temperature scale $T_0$ and parametrized as $A(x,T)= (f/T_0)^2[T-T^*(x)]e^{T/T_0}$, $B=bf^4/T_0^3$ and $C(x)=xcf^2/T_0$~\cite{Banerjee_1}. The quadratic term coefficient $A$ is proportional to $(T-T_{lp})$ where $T_{lp}$ is the local pairing scale temperature and in our theory we identify it to be the psudogap temperature scale $T^{*}$~\cite{Timsuk_1999}. Cooling down from above $T^{*}$, the pairing scale $\langle \Delta_m\rangle$ increases with noticible change in magnitude~\cite{Banerjee_1} while $A$ changes sign. Across the phase diagram $T^{*}$ is considered to be varying with doping concentration $x$ as a simplified linear form $T^{*}(x)=T_0(1-\frac{x}{x_c})$ with $T_0\simeq 400~\rm{K}$ at zero doping and vanishing at a doping concentration $x_c=0.3$. The exponential factor $e^{T/T_0}$ suppresses average local gap magnitude $\langle \Delta_m\rangle$ at high temperatures ($T\apgt T^*(x)$) with respect to its temperature independent equipartition value $\sqrt{T/A(x,T)}$ which will result from the simplified form of the functional (Eq.\eqref{Eq.functional2}) being used over the entire range of temperature. In the range of temperature of our study the role of this factor is not very crucial, for a detailed discussion see ref~\cite{Banerjee_1}. The parameter $B$ is chosen as a doping independent positive number and the form of $C$ is chosen to be proportional to $x$ for small doping. The reason for such a choice can be understood from the Uemura correlations~\cite{Uemura1989} where superfluid density $\rho_s \propto x$ in the underdoped region of the cuprates. Further elaborate details about the functional and coefficients can be found in the appendix of Refs.~\cite{Sarkar_2016,Banerjee_1}.

\section{More on transport currents, coefficients and magnetization:}\label{app.Current_densities}
The Nernst effect is the off-diagonal component of the thermopower tensor $\hat{Q}$, measured in the absence of electrical currents
\begin{equation}
{\bf J}_{tr}= \sigma {\bf E} + \alpha(-\nabla T)
\end{equation}
where $J_{tr}$ is transport current, ${\bf E}$ is the electric field and $\nabla T$ is the temperature gradient.
$\hat{Q}=\hat{\sigma}^{-1}\hat{\alpha}$ is the thermopower tensor.
Here
\begin{equation}
\hat{\sigma}=\begin{pmatrix}
\sigma_{xx} & \sigma_{xy}\\
\sigma_{yx} & \sigma_{yy}
\end{pmatrix}
~~\mathrm{and}~~\hat{\alpha}=
\begin{pmatrix}
\alpha_{xx} & \alpha_{xy}\\
\alpha_{yx} & \alpha_{yy}
\end{pmatrix}
\end{equation}
For an isotropic system, $\sigma_{xx}=\sigma_{yy}$ and $\alpha_{xx}=\alpha_{yy}$. Further, $\sigma_{xy}=-\sigma_{yx}$ and $\alpha_{xy}=-\alpha_{yx}$.
Therefore the thermopower tensor 
\begin{eqnarray}
&& \hat{Q} =  \sigma^{-1}\alpha \\
&&   = \frac{1}{\sigma_{xx}^2+\sigma_{yy}^2}\begin{pmatrix}
\sigma_{xx} & -\sigma_{xy}\\
\sigma_{xy} & \sigma_{xx}
\end{pmatrix} \begin{pmatrix}
\alpha_{xx} & \alpha_{xy}\\
-\alpha_{xy} & \alpha_{xx}
\end{pmatrix}
\end{eqnarray}
The Nernst coefficient
\begin{equation}
Q_{xy}=-Q_{yx}=\frac{\alpha_{xy}\sigma_{xx}-\sigma_{xy}\alpha_{xx}}{\sigma_{xx}^2+\sigma_{xy}^2}=(\frac{\alpha_{xy}}{\sigma_{xx}}-S\tan\Theta_H),
\end{equation}
where $\Theta_H=\tan^{-1}(\frac{\sigma_{xy}}{\sigma_{xx}})$ is the Hall angle and $S(Q_{xx}=Q_{yy})$ is thermopower.

Let ${\bf J}_{\mathrm{tot}}^{e}({\bf r})$, ${\bf J}_{\mathrm{tot}}^{Q}({\bf r})$ and ${\bf J}_{\mathrm{tot}}^{E}({\bf r})$ be the total charge, heat and energy current densities at position ${\bf r}$ in the sample. Each of these current densities is a sum of a transport part and magnetization part. The latter exists even in equilibrium and needs to be subtracted to obtain the transport contributions.
If $\Phi(\bf r)$ is the electric potential at ${\bf r}$, these currents are related to each other as
\begin{equation}\label{JQ.tot}
{\bf J}_{\rm tot}^Q({\bf r})={\bf J}^{E}_{\rm tot}({\bf r})-\Phi({\bf r}){\bf J}_{\rm tot}^e (\bf {r})
\end{equation}
The transport part of the current densities have a similar relation 
\begin{equation}\label{JQ.tr}
{\bf J}_{\rm tr}^Q({\bf r})={\bf J}^{E}_{\rm tr}({\bf r})-\Phi({\bf r}){\bf J}_{\rm tr}^e (\bf {r})
\end{equation}

The charge and energy magnetization densities ${\bf M}^e({\bf r})$ and ${\bf M}^E(\bf {r})$ are related with their respective current counterparts such that~\cite{Cooper_1997}
\begin{eqnarray}
{\bf J}^{e}_{\rm mag}({\bf r}) & = & {\bf \nabla} \times {\bf M}^{e}({\bf r})\\
{\bf J}^{E}_{\rm mag}({\bf r}) & = & {\bf \nabla} \times {\bf M}^{E}({\bf r}) ~.\nonumber
\end{eqnarray}

If the surrounding material is non-magnetic, both ${\bf M}^{\rm{e}}({\bf r})$ and ${\bf M}^{\rm{E}}({\bf r})$ vanish outside the material. Therefore integrating over the sample area $\rm{S_A}$ and averaging
\begin{eqnarray}
{\bf\bar J}^{e}_{\rm tr}=\frac{1}{S_A}\int_{S_A}{\bf J}^{e}_{\rm tr}({\bf r})dS_A & = & \frac{1}{S_A}\int_{S_A}{\bf J}^{e}_{\rm tot}({\bf r})dS_A \\
{\bf\bar J}^{E}_{\rm tr}=\frac{1}{S_A}\int_{S_A}{\bf J}^{E}_{\rm tr}({\bf r})dS_A & = & \frac{1}{S_A}\int_{S_A}{\bf J}^{E}_{\rm tot}({\bf r})dS_A ~. \nonumber
\end{eqnarray}
 Utilizing the above relations and Eq.~\eqref{JQ.tot}, Eq.~\eqref{JQ.tr} we get
\begin{equation}
{\bf\bar J}^{Q}_{\rm tr} = \frac{1}{S_A}\left(\int_{S_A}{\bf
J}^{E}_{\rm tot}({\bf r})dS_A - \int_{S_A}\Phi({\bf r}){\bf
J}^{e}_{\rm tr}({\bf r})dS_A\right) ~.
\end{equation}
and
\begin{equation}
{\bf J}_{\rm tot}^Q({\bf r})={\bf J}^{Q}_{\rm tr}({\bf r})+{\bf J}_{\rm {mag}}^E ({\bf r})-\Phi({\bf r})({\bf \nabla} \times {\bf M}^e)
\end{equation}
Now using the identity ${\bf \nabla} \times \Phi {\bf M}^e={\bf \nabla} \Phi\times{\bf M}^{e} + \Phi({\bf{\nabla}}\times {\bf M}^e)$ reduces to 
\begin{equation}
{\bf J}_{\rm tot}^Q({\bf r})={\bf J}^{Q}_{\rm tr}({\bf r})+ {\bf \nabla} \Phi({\bf r})\times {\bf M}^{e}+{\bf \nabla}\times({\bf M}^E-\Phi({\bf r}) {\bf M}^e)
\end{equation}

We identify and note that there is no heat magnetization density ${\bf M}^Q({\bf r})$ such that ${\bf J}^{Q}_{\mathrm mag}({\bf r})  =  {\bf \nabla} \times {\bf M}^{Q}({\bf r})$. In fact,
\begin{equation}
{\bf J}_{\rm {mag}}^Q({\bf r})={\bf \nabla} \Phi({\bf r})\times {\bf M}^{e}+{\bf \nabla}\times({\bf M}^E-\Phi({\bf r}) {\bf M}^e)
\end{equation}
and therefore
\begin{equation}\label{JQ.tr.bar}
{\bf\bar J}^{Q}_{\rm tr} = \frac{1}{S_A}\int_{S_A}({\bf J}^{Q}_{\rm tot}({\bf r}) -{\bf M}^{e}\times{\bf E}) dS_A
\end{equation}
and for ${\bf M}=M\hat{z}$ and ${\bf E}=E\hat{x}$ we obtain
\begin{equation}
\tilde{\alpha}_{yx}=\frac{\bar{J}^{Q(y)}_{\rm tr}}{E}=\frac{\bar{J}^{Q(y)}_{\rm tot}}{E}-M
\end{equation}

\section{Heat and charge current expressions for continuum and lattice models}\label{app.Heat_charge_expressions}

The expressions of charge and heat current~\cite{Ussishkin,Caroli,UD_1991,A_Schmid}  for a continuum Ginzburg-Landau theory
\begin{eqnarray}
{\bf J}^{e}_{GL}=-i C_0\frac{2\pi}{\Phi_0} \langle\Psi^{*}(\nabla-i\frac{2\pi}{\Phi_0}{\bf A})\Psi\rangle  +\mathrm{c.c.}\\
{\bf J}^{Q}_{GL} =  - C_0 \langle (\frac{\partial}{\partial t}-i\frac{2\pi}{\Phi_0}\Phi)\Psi^{*} \left(\nabla - i\frac{2\pi}{\Phi_0}{\bf A} \right) \Psi \rangle + \mathrm{c.c.}
\end{eqnarray}
with $C_0=\frac{\hbar^{2}}{2m^{*}}$ and $\langle...\rangle$ stands for thermal averages.

For the lattice model given by Eq.~\ref{Eq.functional2} the heat current between sites $m$ and $n$ is obtained taking into account a contribution $\rm{J}^E_{\rm{m}\rightarrow \rm{n}}$ from site $\rm{m}$ to $\rm{n}$ and vice versa and subtracting them out as 
\begin{equation}
\rm{J}^{\rm Q}=\frac{1}{2}({\rm J}^{\rm E}_{\rm{m}\rightarrow \rm{n}}-\rm{J}^{\rm E}_{\rm{n}\rightarrow \rm{m}})+M_z(\bf{E}\times \hat{z})
\end{equation}
where $\rm{J}^{\rm E}_{m\rightarrow n}=-\frac{C}{2}\lbrace \frac{\partial \psi^*_m}{\partial t} \sqrt{\frac{\psi_m}{\psi^*_m}}|\psi_n|e^{i\omega_{m,n}}+c.c.\rbrace$ with $\omega_{m,n}=\phi_m-\phi_n-\int_m^n {\bf A}.d{\bf r}$ is a gauge invariant quantity. The charge current expression is $\rm{J}^{\rm e}=\frac{2\pi}{\Phi_0}C\Delta_m\Delta_n\sin(\phi_m-\phi_n-A_{mn})$

For an $XY$ model described by the Hamiltonian, $\mathcal{H}_{XY}=-J\sum_{<mn>}\cos(\phi_m-\phi_n-A_{mn})$, $J$ being the $XY$ coupling, the heat and charge current expressions~\cite{Podolsky} are

\begin{eqnarray}
\rm{J}^{\rm e}_{\rm{XY}}=J\sin(\phi_m-\phi_n-A_{mn})\\
\rm{J}^{\rm Q}_{\rm{XY}}=-\frac{J}{2}(\dot{\phi}_m+\dot{\phi}_n)\sin(\phi_m-\phi_n-A_{mn})+M_z(\bf{E}\times \hat{z})
\end{eqnarray}

One can verify that the frozen amplitude limit of both charge and heat current expressions of our lattice model reduces to these expressions.

\section*{Effective XY-model}
On the under doped side, where $T^*=T_c^{MF} >> T_c$ we can integrate out the amplitude $\Delta_m$ of the pair degrees of freedom ${\psi_m}$ to obtain an effective action $\mathcal{F}_{XY}$  only in terms of the phase.

\begin{equation}
e^{-\beta \mathcal{F}_{XY}(\lbrace \phi_m\rbrace)}=\frac{\int_{0}^{\infty}\prod_m(\Delta_m d\Delta_m)e^{-\beta \mathcal{F}_{0}(\lbrace\Delta_m\rbrace)} e^{-\beta \mathcal{F}_{1}(\lbrace\Delta_m,\phi_m\rbrace)} }{\int_{0}^{\infty}\prod_m(\Delta_m d\Delta_m) e^{-\beta \mathcal{F}_{0}(\lbrace\Delta_m\rbrace)} }=\langle\exp(-\beta \mathcal{F}_{1} )\rangle_{0}
\end{equation}

In the above, we make use of the cumulant expansion i.e. 
\newline

\begin{equation}
\langle\exp(-\beta \mathcal{F}_{1} )\rangle_{0}=\exp\lbrace-\beta\langle\mathcal{F}_{1}\rangle_{0} + \frac{\beta^2}{2}(\langle\mathcal{F}_{1}^{2}\rangle_{0}-\langle\mathcal{F}_{1}\rangle_{0}^{2})+...\rbrace,
\end{equation}

($\langle...\rangle_{0}$ denotes thermal average obtained using $\mathcal{F}_{0}$ only, to obtain,)

\begin{eqnarray*}
\mathcal{F}_{XY}(\lbrace\phi_m\rbrace)=-C\sum_{<m n>} \langle\Delta_{m}\Delta_{n}\rangle_{0} \cos(\phi_m-\phi_n) \\
-\frac{\beta C^{2}}{2} \sum_{<m n>,<l k>} \cos(\phi_m-\phi_n)\cos(\phi_l-\phi_k)[\langle\Delta_{m}\Delta_{n} \Delta_{l} \Delta_{k}\rangle_{0}-\langle\Delta_m \Delta_n\rangle_{0} \langle\Delta_{l}\Delta_{k}\rangle_{0}] \\+ \mathrm{higher~order~terms}
\end{eqnarray*}

By neglecting the fluctuations of amplitudes and retaining just the first of the above expression, an effective $\mathrm{XY}$ model is obtained, i.e. 

\begin{equation}\label{Eq.EffXY}
\mathcal{F}_{XY}[{\phi_m}]=C\bar{\Delta}^2\sum_{<mn>}\cos(\phi_m-\phi_n)
\end{equation}

with $\bar{\Delta}^2 =\frac{\int_0^\infty\Delta^3 Exp[-\beta(A\Delta^2+\frac{B}{2}\Delta^4)]d\Delta}{\int_0^\infty \Delta Exp[-\beta(A\Delta^2+\frac{B}{2}\Delta^4)]d\Delta}$

\end{document}